\documentclass{article}
\usepackage{graphicx} % Required for inserting images
\usepackage{hyperref}
\usepackage{amsmath,amsfonts,amssymb,amsthm,mathrsfs,tikz-cd, enumerate,mathtools,authblk,amscd,tikz-cd,amscd}
\usetikzlibrary[decorations.markings]
\usetikzlibrary{arrows.meta, positioning}
\DeclarePairedDelimiterX\set[1]\lbrace\rbrace{#1}

\newcommand{\ra}{\rightarrow}

\newcommand{\RR}{{\mathbb R}}

\newcommand{\ZZ}{{\mathbb Z}}

\newcommand{\SA}{{\mathscr A}}

\newcommand{\SU}{{\mathscr U}}

\newcommand{\Cech}{{\v{C}ech}}

\newcommand{\bX}{{\mathfrak X}}

\newcommand{\Ad}{{\rm Ad}}
\newcommand{\coker}{{\rm coker}}

\newcommand{\hgamma}{{\hat\gamma}}

\newcommand{\hB}{{\hat B}}
\newcommand{\hS}{{\hat S}}
\newcommand{\hT}{{\hat T}}
\newcommand{\ha}{{\hat a}}

\newcommand{\bH}{{\mathcal H}}

\newcommand{\KT}{{\mathcal{KS}}}

\allowdisplaybreaks

\title{Higher symmetries, anomalies, and crossed squares in lattice gauge theory}
\author{Anton Kapustin, Lev Spodyneiko}
\affil{\normalsize California Institute of Technology, Pasadena, CA 91125}

\begin{document}

\maketitle

\begin{abstract}
\noindent
    We examine higher-form symmetries of quantum lattice gauge theories through the lens of homotopy theory and operator algebras. We show that in the operator-algebraic approach both higher-form symmetries and 't Hooft anomalies arise from considering restrictions of symmetry transformations to spatial regions. The data of these restrictions are naturally packaged into a higher group. For example, for gauge theories in two spatial dimensions, this information is encoded in a crossed square of groups, which is an algebraic model of a 3-group. In general, we propose that higher groups appear in lattice models and QFT as crossed $n$-cubes of groups via a nonabelian version of the \Cech\ construction. 
\end{abstract}
\pagebreak
\tableofcontents

\section{Introduction}

It is generally accepted by now that symmetries of a QFT in $d$ spatial dimensions are described by a $(d+1)$-group rather than an ordinary group. One piece of evidence in favor of this is the realization that a QFT can have $p$-form global symmetries with $p$ ranging from $0$ to $d$ \cite{gensym}. $0$-form symmetries are ordinary symmetries which act on local operators, while higher-form symmetries act only on "nonlocal" operators. But a $(d+1)$-group is more than a mere collection of $d+1$ groups. First of all, $p$-form symmetries with $p>0$ are necessarily abelian. Second, $0$-form symmetries may act on $p$-form symmetries with $p>0$. Third, there may be a further "interaction" between $p$-form symmetries with different values of $p$ \cite{KapThorn,CorDumInt}. All of these points to a specific mathematical structure from the world of homotopy theory called a $(d+1)$-group. 

A $(d+1)$-group can be defined in several equivalent ways. The most economic definition says that a $(d+1)$-group is a connected homotopy $(d+1)$-type, i.e. the homotopy equivalence class of a topological space $\bX$ whose nontrivial homotopy groups occur in degrees from $1$ to $d+1$. $p$-form symmetry group of a QFT is identified with $\pi_{p+1}(\bX)$. Since homotopy groups of degree greater than $1$ are abelian, this naturally explains why $p$-form symmetries are abelian for $p>0$. From this point of view, an ordinary ($0$-form) symmetry $G$ corresponds to a very simple $\bX$, namely, $\bX=BG$ (the space whose fundamental group is $G$, while all other homotopy groups vanish)\footnote{Here and below we only consider abstract groups. For topological groups, the classifying space $BG$ is defined differently.}. Taking higher symmetry into account corresponds to replacing $BG$ with a more complicated space $\bX$ which is a fibration whose  base is $BG$ and fibers have nontrivial higher homotopy groups. A $p$-form symmetry $G$ corresponds to the homotopy type  of a space $\bX$ such that $\pi_n(\bX)$ is trivial for $n\neq p+1$, while $\pi_{p+1}(\bX)=G$. Such a space is usually denoted $K(G,p+1)$.

The second piece of evidence pointing to the importance of homotopy theory in QFT is related to 't Hooft anomalies of global symmetries, both ordinary and higher-form. Here by 't Hooft anomaly we mean any obstruction to gauging a global symmetry. Ignoring for simplicity diffeomorphism symmetry, an 't Hooft anomaly of a $p$-form symmetry $G$ (with discrete topology) is believed to be described by an element of the cohomology group $H^{d+2}(K(G,p+1),U(1))$. This is usually argued as follows \cite{gensym}. To gauge a symmetry of  QFT on a closed Euclidean $d+1$ manifold $M$, one introduces gauge fields which are $(p+1)$-form gauge fields\footnote{More precisely, for a discrete $G$ they are $(p+1)$-cocycles with values in $G$.} on $M$. If we assume that the gauge variation of the partition function can be canceled by "anomaly inflow" from a Topological Field Theory in one dimension higher, then the problem boils down to classifying topological actions in $d+2$ dimensions which depend on a $(p+1)$-form gauge field for group $G$. With some natural assumptions, such actions are precisely elements of $H^{d+2}(K(G,p+1),U(1))$. 

This argument is not completely satisfactory for several reasons. The assumption that the anomaly can be canceled by a TFT in one dimension higher has never been proved in any generality. The notion of gauging is mathematically well defined only on the classical level. For QFTs not defined by a classical action, the above arguments do not provide any practical method of computing the anomaly. Finally, the argument presupposes that one is dealing with a Euclidean QFT, while the notion of 't Hooft anomaly makes sense for QFTs which lack relativistic invariance, and even for quantum lattice systems. For systems that cannot be Wick-rotated to the Euclidean domain, the connection between 't Hooft anomalies and homotopy theory is far from obvious.

From the point of view of homotopy theory, an element of $H^{d+2}(K(G,p+1),U(1))$ can be interpreted as an obstruction to lifting the identity map $K(G,p+1)\ra K(G,p+1)$ to a map $K(G,p+1)\ra \bX$, where $\bX$ is a fibration with base $K(G,p+1)$ and fiber $K(U(1),d+1)$ \cite{Hatcher}. The obstruction, known as the Postnikov class of the fibration, measures the extent to which $\bX$ fails to be homotopy-equivalent to a trivial fibration (the Cartesian product $K(G,p+1)\times K(U(1),d+1)$. This suggests that 't Hooft anomalies of a $p$-form symmetry $G$ arise because $G$ "interacts" with a $d$-form symmetry $U(1)$. A natural conclusion is that both higher-form symmetries and 't Hooft anomalies (including 't Hooft anomalies of ordinary symmetries) are aspects of the same phenomenon. 

Recently, one of us proposed an operator-algebraic approach to higher-form symmetries and their anomalies in the context of quantum lattice systems \cite{AK} (see also \cite{SK} for a related work). Ref. \cite{AK} attaches a $(d+1)$-group to a quantum lattice system in $d$ spatial dimensions for $d=1,2,$ and shows how 't Hooft anomaly of a 0-form symmetry is encoded in the "interaction" between the 0-form symmetry and the $d$-form symmetry described by the Postnikov class. The $d$-form symmetry is present in any lattice model, its generator being the scalar observable. On the other hand, if the algebra of observables is unconstrained, as is the case for standard spin systems, there are no $p$-form symmetries with $0<p<d$. In this paper, we apply the approach of \cite{AK} to quantum 2+1d gauge theories where observables are constrained by the Gauss law. Consequently, higher symmetries are typically present. Their structure and 't Hooft anomalies depend on the precise form of the Gauss law constraints. We show that all this information is encoded in the $(d+1)$-group introduced in \cite{AK} and reproduce some results known from the Euclidean approach. Our methods can be generalized to higher dimensions, although the corresponding algebraic machinery gets increasingly complicated as $d$ becomes larger.

The content of the paper is as follows. Section 2 is a non-technical overview of our approach. Section 3 is a mathematical preparation. We recall an algebraic model for a 3-group (crossed square) and explain how to extract homotopy groups 
$\pi_1,\pi_2,\pi_3$ and other topological information from it. In particular, every 3-group gives rise to a "twist": a quadratic function $q:\pi_2\ra\pi_3$. In Section 4, we study the "vanilla" $\ZZ_2$ gauge theory without electrically charged matter, which is known to have a non-anomalous 1-form $\ZZ_2$ symmetry. We attach a 3-group to it following the recipe from \cite{AK} and show that the generator of the electric 1-form symmetry gives rise to a distinguished $\ZZ_2$ subgroup of $\pi_2$. The quadratic function $q$ annihilates the generator, indicating that this symmetry is anomaly-free. In Section 5, we perform a similar analysis for a $\ZZ_2$ gauge theory with a modified Gauss law studied in the context of 2d bosonization \cite{ChenKapRad}. Vortex excitations of this theory have fermionic statistics. In the Euclidean approach, this is attributed to an anomalous 1-form $\ZZ_2$ symmetry \cite{GaiottoKap}. We verify the presence of an anomaly in our operator-algebraic framework. We also extend the analysis to $\ZZ_n$ gauge theories with a modified Gauss law, as well as to certain $\ZZ_n\times\ZZ_n$ gauge theories with anomalous 1-form symmetries. In Section 6 we discuss a generalization to higher dimensions. It is known that $(n+1)$-groups can be modeled by crossed $n$-cubes of groups \cite{crossedcubes}. We argue that these gadgets naturally arise from lattice models and QFT and are candidates for higher groups of symmetries. Section 7 contains some concluding remarks. In the Appendix, we explain how to construct a quadratic function $\pi_2\ra\pi_3$ from a crossed square of groups.

We use the following notation. For any group $G$ and any $h\in G$ the automorphism $g\mapsto h g h^{-1}$ is denoted $\Ad_h$. By a commutator we typically mean a commutator in the group sense, i.e. $[g,h]=ghg^{-1}h^{-1}$. Since we deal with nonabelian groups, both the neutral element and the trivial group are denoted $1$. We use this notation even we talk about abelian groups, such as higher homotopy groups of a space. Similarly, bilinear and quadratic functions on an abelian group with values in another abelian group are written multiplicatively. For example, a bilinear function on $A\times A$ with values in $U(1)$ is a function $b$ which satisfies $b(xy,z)=b(x,z)b(y,z)$ and $b(x,yz)=b(x,y)b(x,z)$ for all $x,y,z\in A$. A function $q:A\ra U(1)$ is called quadratic if $b(x,y)=q(xy)/q(x)q(y)$ is bilinear and $q(x^n)=q(x)^{n^2}$ for all $x\in A$.

This work was supported by the Simons Investigator Award.

\section{A non-technical overview}

This section provides an informal overview of the paper's main ideas. For more details, see the following sections and \cite{AK}.

An t' Hooft anomaly is usually understood as an obstruction to gauge a symmetry, where gauging means promoting a global symmetry to a local symmetry. The first step in gauging is writing down the global symmetry acting on the whole space as a product of actions on individual sites. A standard assumption is that localized terms still form a representation of the symmetry group. However, there could be an anomaly that manifests itself in the absence of a consistent way to do so. A well-known example is the Nayak-Else anomaly \cite{NayakElse}, where the localized action can only be a representation of the extension of the symmetry group rather than the group itself.

In this paper, we will take an obstacle to separate the actions of a global symmetry into actions localized on subregions (in such a way that each of them forms a group representation) as the definition of the t'Hooft anomaly. The ultimate objective is to act on individual sites; however, the anomaly manifests itself already when dividing the space into a few regions.

Consider a collection of regions $U_i$ covering the space. To each region $U$ we can assign a group $\KT(U)$ of unitary transformations supported on a finite thickening of $U$. To keep the discussion non-technical, we will not specify exactly what is meant by a unitary transformation localized in a region yet. The above problem can be formulated as a construction of homomorphisms $f_i: G\rightarrow \KT(U_i)$ given a homomorphism $f: G\rightarrow \KT(\bigcup_i U_i)$ such that $f$ is a product of $f_i$. Note that there are ambiguities in the procedure associated to the groups $\KT(U_i\bigcap U_j)$, ambiguities of ambiguities associated to $\KT(U_i\bigcap U_j\bigcap U_k)$, etc. 

The collection of group $\KT(U)$ for all $U$ comes with an additional structure. Whenever $U \subset U'$, one has an inclusion $\KT(U) \rightarrow\KT(U') $ as a subgroup. The subgroup is actually normal since conjugation of a transformation localized on $U \subset U'$ by an element of $\KT(U')$ is again a transformation localized on $U$.  If we have two elements $g\in\KT(U)$ and $ g'\in\KT(U')$, we can construct their commutator, which takes value in $\KT(U\bigcap U')$ since it acts trivially away from $U\bigcap U ' $. This gives a map $\KT(U)\times\KT(U')\rightarrow\KT(U\bigcap U')$. 

All of these data should satisfy a number of consistency conditions. Fortunately, there already exists an algebraic construct known as a crossed $n$-cube that encodes precisely this data. It appeared in the mathematical literature as a tool to relate the homotopy groups of a space with the homotopy groups of regions covering it (i.e. a generalization of Seifert–Van Kampen theorem from $\pi_1$ to $\pi_n$). In particular, an $n$-cube can be related to a homotopy $(n+1)$-type.

On the other hand, as explained in the introduction, a homotopy $n$-type can be used as a definition of an $n$-group, and it is believed that higher groups describe symmetries of QFTs. A natural question is whether the crossed $n$-cube $\KT(U)$ arising in the construction above can be used as a definition of the symmetry $n$-group of a lattice system. In this paper, we continue the work of \cite{AK} and perform several consistency checks, showing that our results reproduce the expected outcomes for lattice gauge theories.

We conclude this section with a couple of comments. First, whenever we restrict a symmetry to act only on a subregion, the resulting transformation no longer preserves the Hamiltonian. In the usual gauging procedure, one introduces gauge fields that transform under the symmetry and modifies the Hamiltonian to achieve invariance under local symmetries. Since we are only interested in t'Hooft anomalies, we do not have to implement this. However, the fact that restricted transformations no longer commute with the Hamiltonian is crucial. Because of it, we will work with groups generated by all unitary transformations. We will refer to them as kinematic symmetries to distinguish them from transformations that preserve the Hamiltonian. They can be thought of as analogs of unitary transformations in quantum mechanics or canonical transformations in classical mechanics. Note that the word "symmetry" here is understood as a geometric symmetry of the set of observables at a fixed time. It is a transformation that preserves the symplectic structure of the phase space in the classical case, the Hermitian structure in ordinary quantum mechanics, and the algebra of local observables in the case of quantum lattice systems. The localizability of the symmetry action characterizes the anomalies and can be diagnosed without specifying the Hamiltonian. 

Second, we will study an infinite lattice system, and we must be careful about what we mean by a symmetry transformation. One cannot implement a kinematic symmetry as a unitary operator, since it acts on an infinite number of sites, and an unrestricted infinite product of Hilbert spaces is ill-defined \cite{vonNeumann_direct}. Instead, a kinematic symmetry must be implemented as a symmetry of the algebra of local operators. On the other hand, if the support of a transformation is finite, we will represent it as a unitary operator. This introduces some subtleties, explained below, resulting in a slight generalization of what was inclusion and commutator above.

Lastly, the discussion of this section focused on 0-form symmetry. It turns out that higher-form symmetry naturally appears in this approach. For example, 1-form symmetries will arise from transformations that are localized on both $U_i$ and $U_j$ but not localized on the intersection $U_i\bigcap U_j$. As we will see later, these are elements of $\pi_2$ of the corresponding homotopy type.  

\section{Crossed squares and homotopy theory}
This section provides an overview of the mathematical results needed in the later sections. First, we discuss the topological data that describe a homotopy type, and then we examine crossed squares, which provide a purely algebraic way to encode the data of a homotopy 3-type. 

\subsection{Postnikov towers}

A useful way to think about homotopy types is via Postnikov systems \cite{Hatcher} (also known as Postnikov towers). For any connected topological space $\bX$ with homotopy groups $\pi_i(\bX)=\pi_i$, one can construct an infinite sequence of spaces $\bX_n$ and maps $p_n: \bX\rightarrow \bX_n$, $f_n: \bX_n\rightarrow \bX_{n-1}$, such that
\begin{itemize}
    \item each $\bX_n$ is a homotopy $n$-type, i.e. $\pi_i(\bX_n)=0$ for $i>n$,
    \item each map $p_n$ induces an isomorphism on homotopy groups $\pi_i(\bX) \rightarrow \pi_i(\bX_n)$ for $i\le n$,
    \item each map $f_n: \bX_n \rightarrow \bX_{n-1}$ is the base projection map of the fibration $K(\pi_n,n)\rightarrow \bX_n \xrightarrow{f_n} \bX_{n-1}$ with the base and fiber being  $\bX_{n-1}$ and the Eilenberg-MacLane space $K(\pi_n,n)$, respectively.
    \item $f_n$ and $p_n$ are compatible in the sense that the diagram below is commutative:
\end{itemize} 

\[
\begin{tikzcd}[column sep=small, row sep=small]
& & K(\pi_n, n) \arrow[d] & \dots & K(\pi_{2}, 2) \arrow[d] & K(\pi_{1}, 1) \arrow[d] \\
& \dots \arrow[r] 
& \bX_n \arrow[r, "f_{n}"] 
& \dots \arrow[r, "f_3"]
& \bX_2 \arrow[r, "f_2"] 
& \bX_1\arrow[r, "f_1"] 
& \bX_0=pt 
\\
& & \bX  \arrow[u,"p_n", dashed]  \arrow[urr,"p_2", dashed,bend right=5] \arrow[urrr, "p_1",dashed,bend right=10]  \arrow[urrrr,"p_0", dashed,dashed,bend right=15] 
\end{tikzcd}
\]

The construction of $\bX_n$ can be visualized as follows. For each non-trivial element of $g\in \pi_{n+1}(\bX)$ we glue an $(n+1)$-ball to $\bX$ along the boundary defined by $g$. This kills  $\pi_{n+1}(\bX)$, while leaving the lower-dimensional homotopy groups unaffected. Proceeding in the same fashion, one can inductively kill all homotopy groups $\pi_i(\widetilde \bX)$ with $i>n$. 

The space $\bX$ is an (inverse) limit of its tower and uniquely defined by it up to homotopy. If $\bX$ is a homotopy  $n$-type, the tower can be terminated at the $n$th term $\bX_n \simeq \bX$. If we think of a connected $n$-homotopy type as an $n$-group, then the space $\bX_i$ is constructed from $\bX$ by forgetting the higher group structure for degrees higher than $i$. 

Each fiber $K(\pi_n, n)$ is a topological group, and the fibrations $K(\pi_n,n)\rightarrow \bX_n \xrightarrow{f_n} \bX_{n-1}$ are principal $K(\pi_n, n)$-bundles\footnote{This is true only in "up to homotopy" sense which is enough for us. Also note that we assume a discrete topology on $\pi_i$ in this paper.  }. Principal $G$-bundles are classified by the classifying space $BG$ in the sense that principal $G$-bundles $G\rightarrow E\rightarrow \bX$  (up to equivalence) are in one-to-one correspondence with the homotopy classes $[\bX, BG]$ of maps $\bX\rightarrow BG$. Thus, each fibration $K(\pi_n,n)\rightarrow \bX_n \xrightarrow{f_n} \bX_{n-1}$ is uniquely determined by an element of $[\bX_{n-1}, BK(\pi_n,n)]\simeq [\bX_{n-1}, K(\pi_n,n+1)]\simeq H^{n+1}(\bX_{n-1},\pi_n)$, where we used the standard facts $BK(\pi_n,n) \simeq K(\pi_n,n+1)$ and  $[\bX, K(G,n)]\simeq H^{n}(\bX,G)$ (see \cite{Hatcher} or any other text book on algebraic topology).

Combining all of the above together, we find that the $n$-group is uniquely determined by its homotopy groups $\pi_i$  and elements of cohomology groups $k_i \in H^{i+1}(\bX_{i-1},\pi_n) $ called Postikov classes for $i=1,\dots,n$. We consider a couple of illustrative examples.

Let all Postnikov classes be trivial. Then each fibration is trivial and the resulting space  $n$-group is just a direct product $\bX = K(\pi_1,1)\times\dots\times K(\pi_n,n) 
$. Physically, it corresponds to the case where different $p$-form symmetries do not interact with each other, and the $n$-group can be thought of as a direct product of usual groups $\pi_i$. The Postnikov classes $k_i$ can be regarded as a measure of how $n$-group structure differs from a direct product.

Next consider the case where $\pi_i=1$ except for $\pi_1$ and $\pi_n=U(1)$. There is only one Postnikov class $k_n \in H^{n+1}(\bX_{n-1},\pi_n) =H^{n+1}(K(\pi_1,1),U(1))$ that can be non-trivial. It is a group cohomology class $H^{n+1}(\pi_1,U(1))$. More generally, when only two homotopy groups are non-trivial, the homotopy type is completely determined by the nontrivial homotopy groups and a single Postnikov class. 
%and can be interpreted as Nayak-Else anomaly \cite{AK}. 
\subsection{Crossed squares}
Homotopy types can also be described algebraically. There are several algebraic models for 3-groups, i.e. connected homotopy types whose only nonvanishing homotopy groups are $\pi_1,\pi_2,\pi_3$. The one most convenient for us is crossed squares of groups, or simply crossed squares \cite{crossedsquare,Conduche2}.

Recall first the notion of a crossed module, which is an algebraic model for a 2-group \cite{Whitehead,BaezLauda}. A crossed module is a pair of groups $H,G$, a homomorphism $\partial:H\ra G$, and an action of $G$ on $H$ by automorphisms. These data must satisfy two conditions:
\begin{align}
    \partial  ({}^g\!h)&=g(\partial h) g^{-1},\ \forall g\in G,\ \forall h\in H,\\
    {}^{\partial h_0}\!h_1&=h_0 h_1 h_0^{-1},\ \forall h_0,h_1\in H.
\end{align}
Here ${}^g\!h\in H$ denotes the result of applying $g\in G$ to $h\in H$. The homotopy groups of the corresponding homotopy 2-type $\bX$ are $\pi_1=\coker\partial$ and $\pi_2=\ker\partial$; the above conditions ensure that $\pi_1$ is a group and $\pi_2$ is an abelian group on which $\pi_1$ acts by automorphisms. One can also extract from the crossed module a class $\beta\in H^3(\pi_1,\pi_2)$. The homotopy type of $\bX$ is completely determined by the triple $(\pi_1,\pi_2,\beta)$. A simple example of a crossed module is given by a triple $(H,G,\partial)$ where $H$ is a normal subgroup of $G$ and $\partial$ is the inclusion map. Then $\pi_1=G/H$, $\pi_2=1$, and $\beta$ is trivial. The corresponding homotopy 2-type is the 1-type $K(G/H,1)$.

A crossed square is a quadruple of groups and group homomorphisms which fit into a commutative diagram:
\begin{align}
\begin{CD}
L @>f>> M \\
@VgVV @VVvV \\
N @>u>> P
\end{CD}
\end{align}
In addition, $P$ acts on $L,M,N$ by automorphisms in such a way that the maps $f$ and $g$ are $P$-equivariant, and we have the following four crossed modules:
\begin{align}
\begin{CD}
M @>v>> P,\ N @>u>> P,\ L @>v\circ f>> P,\ \ L @>u\circ g>> P.
\end{CD}
\end{align}
Finally, part of the data of a crossed square is a map $\eta:M\times N\ra L$ which satisfies 
\begin{align}
\label{eq: f eta}
f(\eta(m,n))&=m\, {}^{u(n)}\!m^{-1}, &g(\eta(m,n))&={}^{v(m)}\!n \, n^{-1},\\
\label{eq: eta boundary}
\eta(f(l),n)&=l\,{}^{u(n)}\!l^{-1},  &\eta(m,g(l))&={}^{v(m)}\!l\, l^{-1},\\
\label{eq: eta exp}
\eta(m m',n)&={}^{v(m)}\!\eta(m',n)\eta(m,n), & \eta(m,nn')&=\eta(m,n)\, {}^{u(n)}\!\eta(m,n'),\\
\label{eq: eta equivariance}
\eta({}^p\!m,{}^p\!n)&={}^p\!\eta(m,n), & &
\end{align}
for all $m,m'\in M$, $n,n'\in N$ and $p\in P$.

To get an example of a crossed square, let $P$ be an arbitrary group, let $M,N$ be normal subgroups of $P$, and set $L=M\cap N$. $P$ acts on $L,M,N$ by conjugation. The homomorphisms that make up the sides of the square are inclusion homomorphisms. The map $\eta$ is given by $\eta(m,n)=m n m^{-1} n^{-1}$. The identities for $\eta$ are straightforward to verify.  

Suppose we are given a crossed square as above. The homotopy groups of the corresponding homotopy 3-type $\mathfrak X$ can be extracted as follows \cite{Conduche2}. Note first that $N$ acts on $M$ via $n: m\mapsto {}^{u(n)}\! m $. Thus, we can form a semidirect product $Q=M\rtimes N$. Second, we define homomorphisms $\delta:L\ra M\rtimes N$ and $\partial:M\rtimes N\ra P$ by
\begin{align}
    \delta:l\mapsto (f(l)^{-1},g(l)),\quad \partial:(m,n)\mapsto v(m)u(n).
\end{align}
One can easily check that both $\delta$ and $\partial$ are group homomorphisms. Also, $\partial\circ\delta$ is a trivial homomorphism, i.e. it sends all of $L$ to $1\in P$. Thus, we get a complex of nonabelian groups 
\begin{align}
    \begin{CD}
L @>\delta>> M\rtimes N @>\partial>> P
\end{CD}
\end{align}
It can be checked that the images of both $\partial$ and $\delta$ are normal subgroups, so that the homology sets of the complex are groups. These homology groups are the homotopy groups of the 3-group  $\bX$:
\begin{align}
    \pi_1(\bX)=\coker\partial,\quad \pi_2(\bX)=\ker\partial/{\rm im}\,\delta,\quad \pi_3(\bX)=\ker\delta.
\end{align}

To compute the homotopy groups, one does not need to know the function $\eta$. But it is needed to determine other data of a homotopy 3-type. In particular, it allows one to define a function $q:\pi_2\ra\pi_3$. From the topological point of view, this function arises by composing a map $S^2\ra \bX$ representing $x\in\pi_2(\bX)$ with the Hopf fibration map $S^3\ra S^2$. To compute this function algebraically, we first define a map $\{\cdot,\cdot\}:Q\times Q\ra L$ by letting 
\begin{align}
    \{(m,n),(m',n')\}=\eta(m,n n' n^{-1})^{-1}.
\end{align}
Then for any $x\in \ker\partial$ we let
\begin{align}
    q(x)=\{x,x\}.
\end{align}
It is shown in the Appendix that $q(x)$ depends only on the equivalence class of $x$ in $\pi_2=\ker\partial/{\rm im}\, \delta$, takes values in $\pi_3=\ker\delta$, and that it is a quadratic function of $x$ equivariant under the action of $\pi_1$ on $\pi_2,\pi_3$.

\section{$\ZZ_2$ gauge theory in two dimensions}

\subsection{Operator-algebraic basics}

In the Hilbert space approach to quantum mechanics, an observable is a bounded operator in the Hilbert space $\bH$ of the system. Observables form an associative algebra of bounded operators $\SA=B(\bH)$. It is equipped with an antilinear map  $a\mapsto a^*$ such that $(ab)^*=b^*a^*$ for all $a,b\in\SA$. The resulting mathematical structure is called a $*$-algebra (with a unit). A special role is played by unitary observables and self-adjoint observables. In the operator-algebraic approach, observables are elements of an abstract $*$-algebra $\SA$ with a unit. In the context of quantum lattice systems, $\SA$ is either an infinite product of on-site algebras or, in the case of lattice gauge theories, a sub-algebra of such an infinite product defined by Gauss law constraints. As in the Hilbert space approach, a special role is played by unitary and self-adjoint observables. For example, a unitary observable is defined as an element $u\in\SA$ such that $uu^*=u^*u=1.$

In the Hilbert space approach, symmetries are described by unitary or anti-unitary operators on the Hilbert space. This is the statement of Wigner`s theorem \cite{Wigner}. In the operator-algebraic approach, the analog of a unitary transformation is an automorphism of $\SA$, i.e. an invertible linear map $\alpha:\SA\ra\SA$ that satisfies $\alpha(ab)=\alpha(a)\alpha(b)$, $\alpha(a)^*=\alpha(a^*)$ for all $a,b\in\SA$. Automorphisms of a $*$-algebra form a group. We will call it the group of kinematic symmetries, since it contains all transformations that preserve algebraic relations between observables at a fixed time but do not necessarily commute with the dynamics. Every unitary observable $u$ defines an automorphism $\Ad_u$ that sends $a\in\SA$ to $uau^{-1}$. Importantly, unlike in the Hilbert space approach, for infinite-volume systems physically interesting automorphisms of $\SA$ often do not have this form.

In the Hilbert space approach, dynamics is described by a group homomorphism from the additive group of real numbers $\RR$ to the unitary group of $\bH$, i.e. a one-parameter family of unitary operators $u_t$, $t\in\RR$, such that $u_0=1$ and $u_t u_s=u_{t+s}$ for all $s,t\in\RR$. A symmetry of dynamics is a unitary operator that commutes with $u_t$ for all $t$. In the operator-algebraic approach, dynamics is described by a one-parameter family of automorphisms $\alpha_t$, $t\in\RR$, satisfying $\alpha_0=1$, $\alpha_t\circ\alpha_s=\alpha_{s+t}$ for all $s,t\in\RR$. A symmetry of dynamics is an automorphism of $\SA$ which commutes with all $\alpha_t$. Such automorphisms form a group which one might call the group of dynamic symmetries. It is a subgroup of the group of kinematic symmetries. 

Our intention is to promote the group of dynamic symmetries of a lattice gauge theory to a higher group. This is possible thanks to an additional structure present in lattice systems and QFT: the ability to restrict automorphisms to regions, i.e. the ability to replace an automorphism with a different automorphism which acts in the same way deep inside the region and acts trivially sufficiently far from the region. In the case of symmetries acting on-site, there is a natural way to do this: acting only on the sites inside the region of interest. However, restricting a dynamical symmetry to a region leads only to a kinematic symmetry. This is one of the reasons to consider kinematic symmetries. More generally, one may consider automorphisms generated by (finite-depth quantum) circuits. Circuits can be restricted to regions as well, by dropping gates that are localized far from the region of interest. However, this procedure is non-unique. Homotopy theory helps one to extract invariant information despite this non-uniqueness. 

Every circuit generates an automorphism, but different circuits can generate the same automorphism. In \cite{AK}, it was important to distinguish circuits from the automorphisms they generate. In this paper, we use the results of \cite{AK} as a black box and will be sloppier with terminology. 

\subsection{Kinematic symmetries of the $\ZZ_2$ gauge theory}

Consider a quantum $\ZZ_2$ gauge theory on an infinite square lattice in $\RR^2$. In such a theory, there is a qubit on every edge $e$ described by Pauli matrices $X_e,Y_e,Z_e$, and a Gauss law constraint for every vertex $v$. The constraint associated to $v$ says that physical observables are polynomials in the Pauli matrices  which commute with 
\begin{align}
    A_v=\prod_{e\supset v} X_e
\end{align}
for all vertices $v$. A graphic representation of $A_v$ is shown in Fig. 1. We will call polynomials commuting with all $A_v$ gauge-invariant observables. They form a $*$-algebra $\SA$. It is easy to see that a general gauge-invariant observable is a polynomial in $X_e$ and $B_p=\prod_{e\subset p} Z_e$, where $p$ is a plaquette (elementary square of the lattice).\footnote{With periodic boundary conditions, the algebra $\SA$ also contains topologically nontrivial Wilson loops, but in infinite geometry they do not exist, while infinite Wilson loops are not in the algebra $\SA$.} Unitary gauge-invariant observables are those which satisfy in addition $a^*a=aa^*=1$. They form a group $\SU\subset\SA$. The usual Hamiltonian is a formal sum of gauge-invariant observables:
\begin{align}\label{eq:ham}
H=-g^2\sum_{e} X_e-\frac{1}{g^2}\sum_p B_p
\end{align}
From a mathematical viewpoint, a Hamiltonian is a derivation of $\SA$ that can be exponentiated to a time-evolution automorphism. That is, while the sum defining $H$ is formal, the expression $[H,a]$ is a well-defined element of $\SA$ for any $a\in \SA$.

\begin{figure}
\centering
\begin{tikzpicture}[
    scale=1.5,
    normal_edge/.style={gray, line width=1pt},
    red_edge/.style={red, line width=3pt},
    vertex/.style={circle, fill=black, inner sep=1.5pt}
]

% Draw all horizontal edges in gray
\foreach \x in {0,1,2,3} {
    \foreach \y in {0,1,2,3,4} {
        \draw[normal_edge] (\x,\y) -- (\x+1,\y);
    }
}

% Draw all vertical edges in gray
\foreach \x in {0,1,2,3,4} {
    \foreach \y in {0,1,2,3} {
        \draw[normal_edge] (\x,\y) -- (\x,\y+1);
    }
}

% Draw the four red edges that share a vertex (choosing vertex at (2,2))
% These will overdraw the gray edges
\draw[red_edge] (2,2) -- (2,3);   % up
\draw[red_edge] (2,2) -- (2,1);   % down
\draw[red_edge] (2,2) -- (1,2);   % left
\draw[red_edge] (2,2) -- (3,2);   % right

% Draw all vertices
\foreach \x in {0,1,2,3,4} {
    \foreach \y in {0,1,2,3,4} {
        \node[vertex] at (\x,\y) {};
    }
}

% Highlight the central vertex where red edges meet
\node[vertex, fill=red] at (2,2) {};

\end{tikzpicture}
\caption{A graphic representation of $A_v$ for a particular $v$. Red edges denote Pauli $X$ generators.} 
\end{figure}
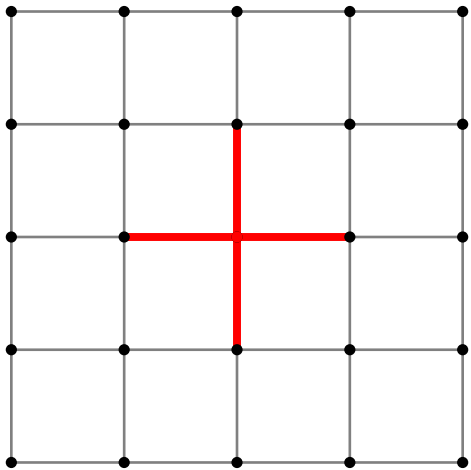

Generally speaking, symmetries are automorphisms of the algebra of observables.  We define kinematic symmetries of this system as arbitrary automorphisms generated by gauge-invariant circuits, i.e. circuits made of unitary gauge-invariant observables. Kinematic symmetries need not preserve the Hamiltonian. 

Following \cite{AK}, one can assign a crossed square of kinematic symmetries to this system as follows. First, we define an $r$-thickening of a region $U\subset\RR^2$ as the set of all points of $\RR^2$ that are within distance $r$ from $U$. We say that a circuit is approximately localized in a region $U$ if all its gates are localized within some thickening of $U$. Pick a half-plane. By a choice of coordinates, we may assume it is given by $y<0$ (lower half-plane). Let $P$ be the group of all gauge-invariant circuits that are approximately localized in the lower half-plane. Let $M$ (resp. $N$) be the group of all gauge-invariant circuits which are approximately localized on the left (resp. right) half of the $x$-axis. Let $L$ be the group of unitary gauge-invariant observables $\SU$. $M,N$ can be regarded as normal subgroups of $P$, and  $P$ acts on $L,M,N$ in the obvious way. Next, we define the homomorphisms between the vertices of the square. The homomorphisms $M\ra P$ and $N\ra P$ are subgroup inclusions. The homomorphisms $L\ra M$ and $L\ra N$ are given by $u\mapsto\Ad_u$.  

Finally, we need to specify the map $\eta:M\times N\ra L$. To any two circuit-generated automorphisms $m,n$ approximately localized on the left and right half-lines, one can attach their "commutator"  taking values in $\SU$ \cite{AK}. We put the word "commutator" in quotes because the true commutator takes values not in $\SU$, but in automorphisms approximately localized at the origin, i.e. in the image of $\SU$ in the group of automorphisms under the map $u\mapsto \Ad_u$. This map has a nontrivial kernel, which consists of multiples of the identity observable. It is a somewhat nontrivial fact that one can canonically uplift the true commutator to an element of $\SU$ \cite{AK}. To compute the "commutator", one first truncates each of the two circuit-generated automorphisms $m,n$ to sufficiently large finite regions. The truncated automorphisms $\tilde m,\tilde n$ have the form $\Ad_{u},\Ad_v$, respectively, where $u,v$ are some unitary observables. One defines the "commutator" of $m,n$ to be $uvu^{-1}v^{-1}$. It is shown in \cite{AK} that this definition is independent of any choices made. We will denote the "commutator" of $m\in M$ and $n\in N$ by $\eta(m,n)$. It follows from the results of \cite{AK} that the above data define a crossed square. We will refer to it as the kinematic crossed square.

Let us discuss the homotopy groups of the kinematic crossed square. We denote $Q=M \rtimes N$ and let $\delta:L\ra Q$ and $\partial:Q\ra P$ be the homomorphisms defined as in Section 2. We clearly have $\pi_3=\ker\delta=U(1)$. We will now show that $\pi_1=\coker\partial$ is non-trivial and infinite.  To see this, note that for any sequence of plaquettes $\{p_n\}_{n\in\ZZ}$, we can define an automorphism which acts as follows: it flips the sign of all $B_{p_n}$ and leaves invariant all other $B_p$. It also leaves invariant all $X_e$. Now, suppose the sequence of $y$-coordinates of these plaquettes diverges as $n\ra\infty$. In that case, it is clear that the corresponding automorphism cannot lie in the image of $\partial:Q\ra P$, therefore is a nontrivial element of $\pi_1$. By varying the sequence of plaquettes, one can easily see that $\pi_1$ is infinite. 

Finally, we claim that $\pi_2$ has a distinguished $\ZZ_2$ subgroup  and thus is nontrivial. To see this, let $x=(m,n)\in Q$. If $\partial x=mn=1$, the automorphisms $m$ and $n=m^{-1}$ must act trivially outside of some square centered at $0$. Let us take $m$ to be the automorphism as above, which flips a single $B_p$. It clearly squares to identity and lies in the intersection of $M$ and $N$. Thus if we let $x=(m,m)\in Q$, then $\partial x=1$. $x$ does not lie in the image of $\delta:L\ra Q$. Indeed, if this automorphism had the form $\Ad_u$ for some gauge-invariant $u$, then $u$ would have to be a polynomial function of $X_e$ only. But it is easy to see that no finite collection of $X_e$ could flip the sign of just a single $B_p$. Finally, it is easy to check that the class of the automorphism in $\pi_2$ does not depend on the choice of the flipped plaquette.

The nontrivial element of $\pi_2$ constructed above can also be described in terms of "string automorphisms". 
Pick any path $\hgamma$ on the dual lattice that stays in the lower half-plane. One can attach to $\gamma$ an automorphism $\mu^X_\hgamma=\Ad_{X_\hgamma}$, where $X_\hgamma$ is a formal product $\prod_{e\cap\hgamma\neq\emptyset} X_e$. $\mu^X_\hgamma$ leaves all $X_e$ invariant and flips the sign of $B_p$ when $p$ is the endpoint of $\hgamma$ (all other $B_p$ are invariant). Thus $\mu^X_\hgamma$ depends only on the endpoint of $\hgamma$ and coincides with the automorphisms discussed in the previous paragraph. The generator of $\ZZ_2\subset\pi_2$ can be taken to be $(\mu^X_{\hgamma_-},\mu^X_{\hgamma_+})\in M\rtimes N$, where $\hgamma_+$ and $\hgamma_-$ are paths which run along the $x$-axis from $0$ to $x=+\infty$ and $x=-\infty$, respectively. 

In the context of the toric code \cite{Kitaev_toric}, the automorphisms $\mu^X_\gamma$ are known as "magnetic string operators". When they act on the ground state of the toric code, they create magnetic excitations localized at the endpoint of $\hgamma$. These excitations are eigenstates of the toric code Hamiltonian. This follows from
\begin{align}\label{eq:muH}
\mu^X_\gamma(H_{toric})=H_{toric}+2t B_{p_0},
\end{align}
where $p_0$ is the endpoint of $\hgamma$ 
and the fact that for any plaquette $p$ the operator $B_p$ commutes with the toric code Hamiltonian
\begin{align}
    H_{toric}=-s\sum_v A_v-t \sum_p B_p .
\end{align}
We will refer to $\mu^X_{\hgamma}$ as $X$-string automorphisms. Unlike in the toric code case, the excitation created by $\mu^X_\hgamma$  is not an eigenstate of the Hamiltonian (\ref{eq:ham}).

\subsection{Dynamic symmetries of the $\ZZ_2$ gauge theory}

Let us now bring in the Hamiltonian in the picture. Ordinarily, a dynamic symmetry is required to preserve the Hamiltonian. In the context of lattice systems, a dynamic symmetry is a circuit-generated automorphism $\alpha$ of $\SA$ such that $\alpha(H)=H$. In our approach, we want to regard a 0-form symmetry as an element of $\pi_1$ of the crossed square defined in the previous section. The corresponding circuit is obtained by truncating $\alpha$ to a half-plane. The precise recipe for truncation is not essential since it does not affect the equivalence class of the circuit in $\pi_1$. Any such truncation preserves the Hamiltonian far away from the boundary of the half-plane. 

This motivates us to define a dynamic 0-form symmetry as an element of $\pi_1$ that preserves $H$ up to a derivation approximately localized on the line $y=0$. Together, such elements form a subgroup $\pi^H_1$ of $\pi_1$. Further, we can define $P^H$ as the subgroup of $P$ which consists of gauge-invariant circuits that preserve $H$ up to a derivation approximately localized on the line $y=0$. $P^H$ along with $M,N,L$ forms a crossed square which we call the dynamic crossed square. Its homotopy groups are $\pi^H_1,\pi_2,\pi_3$. The dynamic crossed square defines a 3-group of dynamic symmetries $\bX^H$. Note that with this definition, $p$-form symmetry groups with $p>0$ are independent of the Hamiltonian. This is in accord with the lore that higher-form symmetries are more robust than 0-form symmetries. In the case of the $\ZZ_2$ gauge theory, $\pi^H_1$ does not have any obvious nontrivial elements, so below we will focus on 1-form symmetries.

\subsection{Anomalies}

Suppose a group $G$ acts on a 2d lattice system by circuit-generated automorphisms that preserve the Hamiltonian. As explained in the previous section, truncating this action to a half-plane gives rise to a homomorphism $\rho:G\ra\pi^H_1$. It was proposed in \cite{AK} to define 't Hooft anomaly as an obstruction to finding a map of 3-groups $BG\ra \bX^H$ which induces a given homomorphism $\rho:\pi_1(BG)\ra \pi^H_1$. The idea is that the 3-group $\bX^H$ describes localizable symmetries of the system, so finding such a map is equivalent to saying that the action of $G$ is localizable. Conversely, an obstruction to finding such a map means that the $G$-action is not localizable, which we take as a synonym for "not gaugeable". This gives a concrete prescription for computing the anomaly, which agrees with the physics lore. For example, homotopy theory tells us that the first obstruction is the Postnikov class in $H^3(G,\pi_2)$. The relation of this class to  't Hooft anomalies is well known from the study of symmetries of 2+1d TQFTs \cite{Gcrossed}, where the role of $\pi_2$ is played by the group of abelian anyons. 

Concretely, the anomaly can be computed as follows. One forms a crossed module $Q/{\rm im\,}\delta\ra P$, computes the associated Postnikov class following the standard prescription (see \cite{NayakElse,AK} for details) and then pulls it back to $G$ via $\rho$

In the case of the vanilla $\ZZ_2$ gauge theory, there are no obvious 0-form symmetries, but there is a 1-form $\ZZ_2$ symmetry, i.e. a homomorphism $\ZZ_2\ra \pi_2$ whose generator is given by the $X$-string automorphisms. More generally, it is natural to define an action of an abelian group $G$ by 1-form symmetries as a homomorphism $\rho:G\ra  \pi_2$ and define the 't Hooft anomaly of this action as an obstruction to finding a map of 3-groups $K(G,2)\ra \bX^H$ which induces a given homomorphism $\rho:\pi_2(K(G,2))\ra \pi_2$. Homotopy theory tells us that the only obstruction is the  Postnikov class in $H^4(K(G,2),\pi_3)=H^4(K(G,2),U(1))$, or equivalently a quadratic function from $G$ to $U(1)$. This is in agreement with expectations from the Euclidean approach \cite{gensym}.

Concretely, the 't Hooft anomaly of a 1-form symmetry $G$ can be evaluated by composing the quadratic function $q:\pi_2\ra \pi_3$ with the homomorphism $\rho$. Note that the function $q$ is determined by kinematic considerations alone, i.e. it can be computed without knowing the Hamiltonian. This explains the robustness of the anomaly.

In the case of the vanilla $\ZZ_2$ gauge theory, the anomaly of the 1-form $\ZZ_2$ symmetry vanishes. Indeed, let $\hgamma_+$ and $\hgamma_-$ be paths on the dual lattice which begin at $0$ and run along the positive and negative $x$-axis, respectively. The generator of $\ZZ_2$ 1-form symmetry is $x=(\mu^X_{\hgamma_+},\mu^X_{\hgamma_-})$. It is easy to see that the "commutator" of $\mu^X_{\hgamma_+}$ and $\mu^X_{\hgamma_-}$ is trivial, hence $q(x)=\{x,x\}=1$.

In the next section, we give an example of a gauge theory where the 't Hooft anomaly is nontrivial.

\section{Gauge theories with modified Gauss laws}

\subsection{Twisted $\ZZ_2$ gauge theory}

Ref. \cite{ChenKapRad} studied a $\ZZ_2$ lattice gauge theory which is equivalent to a theory of fermions. In the Hamiltonian approach on a square spatial lattice, this theory is defined as follows \cite{ChenKapRad}. The variables are the same as for the vanilla gauge theory, but the Gauss law constraint is modified. Namely, $A_v$ is replaced with
\begin{align}
    A'_v=B_{NE(v)}\prod_{e\supset v} X_e ,
\end{align}
where $NE(v)$ is the plaquette which is north-east of the vertex $v$, see Fig. 2. All these constraints still commute. 

\begin{figure}
\centering
\begin{tikzpicture}[
    scale=1.5,
    normal_edge/.style={gray, line width=1pt},
    red_edge/.style={red, line width=3pt},
    vertex/.style={circle, fill=black, inner sep=1.5pt}
]

% Draw all horizontal edges in gray
\foreach \x in {0,1,2,3} {
    \foreach \y in {0,1,2,3,4} {
        \draw[normal_edge] (\x,\y) -- (\x+1,\y);
    }
}

% Draw all vertical edges in gray
\foreach \x in {0,1,2,3,4} {
    \foreach \y in {0,1,2,3} {
        \draw[normal_edge] (\x,\y) -- (\x,\y+1);
    }
}

% Draw the four red edges that share a vertex (choosing vertex at (2,2))
% These will overdraw the gray edges
\draw[red_edge] (2,2) -- (2,3);   % up
\draw[red_edge] (2,2) -- (2,1);   % down
\draw[red_edge] (2,2) -- (1,2);   % left
\draw[red_edge] (2,2) -- (3,2);   % right

% Draw all vertices
\foreach \x in {0,1,2,3,4} {
    \foreach \y in {0,1,2,3,4} {
        \node[vertex] at (\x,\y) {};
    }
}
% Highlight the central vertex where red edges meet
\node[vertex, fill=red] at (2,2) {};

% Add blue square centered at the northeast lattice square
% The northeast square from vertex (2,2) has center at (2.5,2.5)
% Make it slightly smaller than the unit square
\draw[blue, line width=3pt] (2.15,2.15) rectangle (2.85,2.85);

\end{tikzpicture}

\caption{A graphic representation of $A'_v$. The blue square denotes $B_{NE(v)}$.}
\end{figure}
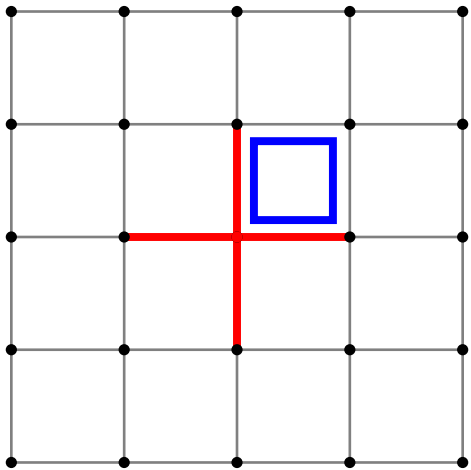

%\end{document}

Gauge-invariant local observables are polynomials in the Pauli generators that commute with all $A'_v$. This is an algebra generated by $B_p$ for all $p$ and $U_e=X_e Z_{r(e)}$ for all $e$. Here $r$ is a map from the set of edges to itself whose definition for horizontal and vertical edges is shown in Fig. 3. The observable $U_e$ for horizontal and vertical edges is shown in Fig. 4. 

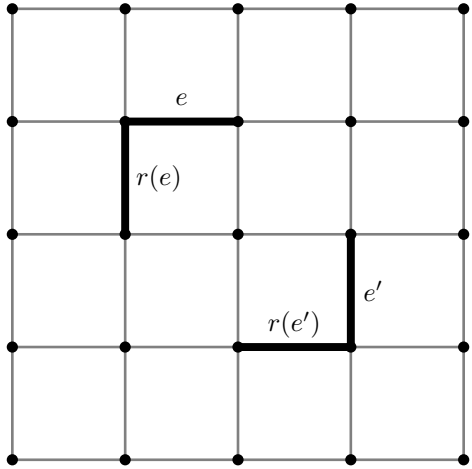
\begin{figure}
\centering
\begin{tikzpicture}[
    scale=1.5,
    normal_edge/.style={gray, line width=1pt},
    red_edge/.style={red, line width=3pt},
    black_edge/.style={black, line width=3pt},
    vertex/.style={circle, fill=black, inner sep=1.5pt}
]

% Draw all horizontal edges in gray
\foreach \x in {0,1,2,3} {
    \foreach \y in {0,1,2,3,4} {
        \draw[normal_edge] (\x,\y) -- (\x+1,\y);
    }
}

% Draw all vertical edges in gray
\foreach \x in {0,1,2,3,4} {
    \foreach \y in {0,1,2,3} {
        \draw[normal_edge] (\x,\y) -- (\x,\y+1);
    }
}

% Draw the four red edges that share a vertex (choosing vertex at (2,2))
% These will overdraw the gray edges
\draw[black_edge] (1,3) -- (2,3);   % e hor
\draw[black_edge] (1,3) -- (1,2);   % r(e) 
\draw[black_edge] (3,2) -- (3,1);   % left
\draw[black_edge] (3,1) -- (2,1);   % right

\node[] at (1.5,3.2) {$e$};
\node[] at (1.3,2.5) {$r(e)$};
\node[] at (3.2,1.5) {$e'$};
\node[] at (2.5,1.2) {$r(e')$};

% Draw all vertices
\foreach \x in {0,1,2,3,4} {
    \foreach \y in {0,1,2,3,4} {
        \node[vertex] at (\x,\y) {};
    }
}
\end{tikzpicture}
\caption{The definition of the map $r$ for horizontal and vertical edges.}
\end{figure}

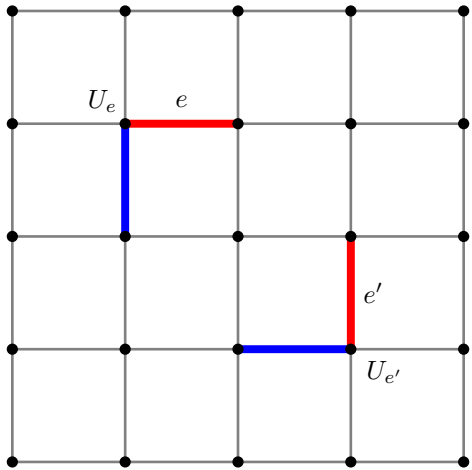
\begin{figure}
\centering
\begin{tikzpicture}[
    scale=1.5,
    normal_edge/.style={gray, line width=1pt},
    red_edge/.style={red, line width=3pt},
    blue_edge/.style={blue, line width=3pt},
    vertex/.style={circle, fill=black, inner sep=1.5pt}
]

% Draw all horizontal edges in gray
\foreach \x in {0,1,2,3} {
    \foreach \y in {0,1,2,3,4} {
        \draw[normal_edge] (\x,\y) -- (\x+1,\y);
    }
}

% Draw all vertical edges in gray
\foreach \x in {0,1,2,3,4} {
    \foreach \y in {0,1,2,3} {
        \draw[normal_edge] (\x,\y) -- (\x,\y+1);
    }
}

% Draw the four red edges that share a vertex (choosing vertex at (2,2))
% These will overdraw the gray edges
\draw[red_edge] (1,3) -- (2,3);   % e hor
\draw[blue_edge] (1,3) -- (1,2);   % r(e) 
\draw[red_edge] (3,2) -- (3,1);   % e vert
\draw[blue_edge] (3,1) -- (2,1);   % r(e)

\node[] at (1.5,3.2) {$e$};
\node[] at (0.8,3.2) {$U_e$};
\node[] at (3.2,1.5) {$e'$};
\node[] at (3.3,0.8) {$U_{e'}$};

% Draw all vertices
\foreach \x in {0,1,2,3,4} {
    \foreach \y in {0,1,2,3,4} {
        \node[vertex] at (\x,\y) {};
    }
}
\end{tikzpicture}
\caption{The definition of gauge-invariant observables $U_e$. Red edges are Pauli $X$, blue edges are Pauli $Z$.}
\end{figure}

The definition of the kinematic and dynamic crossed squares is the same as before. Let us now exhibit a $\ZZ_2$ subgroup of $\pi_2$ and compute its anomaly. We look for an automorphism $m\in M\cap N$  such that $m$ cannot be written as a conjugation by a gauge-invariant local observable. We observe that the circuit shown in Fig. 5 is made of gauge-invariant gates and acts trivially on gauge-invariant observables. Hence, if we decompose it into two gauge-invariant circuits, one of which is supported on the left half-line and the other on the right half-line, the automorphism generated by the left-localized circuit is a candidate for $m$. See Fig. 6. Note that $m^2=1.$ Thus the same automorphism can also be generated by the right-localized circuit shown in Fig. 7. 

\begin{figure}
\centering
\begin{tikzpicture}[
    scale=1.5,
    normal_edge/.style={gray, line width=1pt},
    red_edge/.style={red, line width=3pt},
    blue_edge/.style={blue, line width=3pt},
    vertex/.style={circle, fill=black, inner sep=1.5pt}
]

% Draw all horizontal edges in gray
\foreach \x in {0,1,2,3} {
    \foreach \y in {0,1,2,3,4,5} {
        \draw[normal_edge] (\x,\y) -- (\x+1,\y);
    }
}

% Draw all vertical edges in gray
\foreach \x in {0,1,2,3,4} {
    \foreach \y in {0,1,2,3,4} {
        \draw[normal_edge] (\x,\y) -- (\x,\y+1);
    }
}

\draw[red_edge] (0,3) -- (0,4); 
\draw[red_edge] (1,3) -- (1,4); 
\draw[red_edge] (2,3) -- (2,4); 
\draw[red_edge] (3,3) -- (3,4);  
\draw[red_edge] (4,3) -- (4,4);   

\draw[blue_edge] (1,4) -- (0,4); 
\draw[blue_edge] (2,4) -- (1,4); 
\draw[blue_edge] (3,4) -- (2,4); 
\draw[blue_edge] (4,4) -- (3,4);

\draw[red_edge] (0,2) -- (0,1); 
\draw[red_edge] (1,2) -- (1,1);   
\draw[blue_edge] (1,1) -- (0,1);

\draw[red_edge] (2,2) -- (2,1);   
\draw[blue_edge] (2,1) -- (1,1);
\draw[red_edge] (3,2) -- (3,1);   
\draw[blue_edge] (3,1) -- (2,1);  
\draw[red_edge] (4,2) -- (4,1);   
\draw[blue_edge] (4,1) -- (3,1);

\draw[blue, line width=3pt] (0.15,1.15) rectangle (0.85,1.85);
\draw[blue, line width=3pt] (1.15,1.15) rectangle (1.85,1.85);
\draw[blue, line width=3pt] (2.15,1.15) rectangle (2.85,1.85);
\draw[blue, line width=3pt] (3.15,1.15) rectangle (3.85,1.85);
% Draw all vertices
\foreach \x in {0,1,2,3,4} {
    \foreach \y in {0,1,2,3,4,5} {
        \node[vertex] at (\x,\y) {};
    }
}
\end{tikzpicture}
\caption{Two representations of a circuit in the twisted $\ZZ_2$ theory which acts by the identity automorphism on gauge-invariant observables. The bottom representation is manifestly gauge-invariant.}
\end{figure}
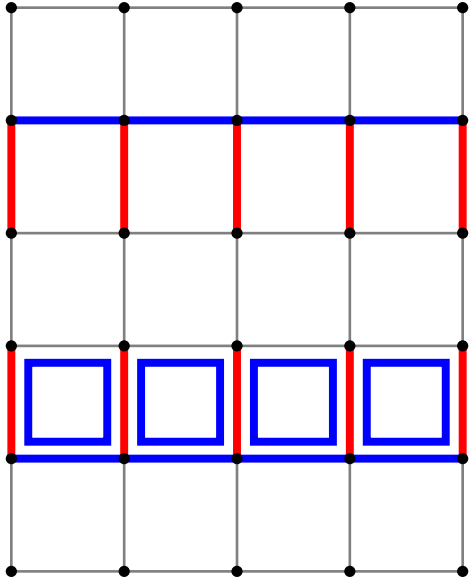

\begin{figure}
\centering
\begin{tikzpicture}[
    scale=1.5,
    normal_edge/.style={gray, line width=1pt},
    red_edge/.style={red, line width=3pt},
    blue_edge/.style={blue, line width=3pt},
    vertex/.style={circle, fill=black, inner sep=1.5pt}
]

% Draw all horizontal edges in gray
\foreach \x in {0,1,2,3} {
    \foreach \y in {0,1,2,3,4,5} {
        \draw[normal_edge] (\x,\y) -- (\x+1,\y);
    }
}

% Draw all vertical edges in gray
\foreach \x in {0,1,2,3,4} {
    \foreach \y in {0,1,2,3,4} {
        \draw[normal_edge] (\x,\y) -- (\x,\y+1);
    }
}

\draw[red_edge] (0,3) -- (0,4); 
\draw[red_edge] (1,3) -- (1,4); 
\draw[red_edge] (2,3) -- (2,4);

\draw[blue_edge] (1,4) -- (0,4);
\draw[blue_edge] (2,4) -- (1,4);
\draw[blue_edge] (1.9,3) -- (1.9,4);

\draw[red_edge] (0,2) -- (0,1); 
\draw[red_edge] (1,2) -- (1,1);   
\draw[blue_edge] (1,1) -- (0,1);

\draw[red_edge] (2,2) -- (2,1);   
\draw[blue_edge] (2,1) -- (1,1);

\draw[blue, line width=3pt] (0.15,1.15) rectangle (0.85,1.85);
\draw[blue, line width=3pt] (1.15,1.15) rectangle (1.85,1.85);

% Draw all vertices
\foreach \x in {0,1,2,3,4} {
    \foreach \y in {0,1,2,3,4} {
        \node[vertex] at (\x,\y) {};
    }
}
\end{tikzpicture}
\caption{Two representations of a string automorphism $m\in M\cap N$ by a circuit supported on the left half-line. The bottom one is manifestly gauge-invariant.}
\end{figure}
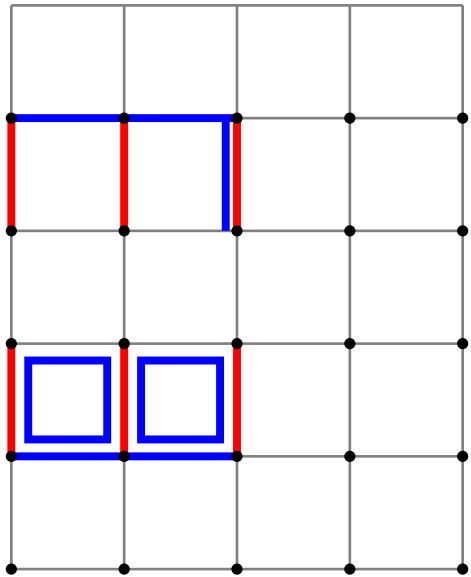

\begin{figure}
\centering
\begin{tikzpicture}[
    scale=1.5,
    normal_edge/.style={gray, line width=1pt},
    red_edge/.style={red, line width=3pt},
    blue_edge/.style={blue, line width=3pt},
    vertex/.style={circle, fill=black, inner sep=1.5pt}
]

% Draw all horizontal edges in gray
\foreach \x in {0,1,2,3} {
    \foreach \y in {0,1,2,3,4,5} {
        \draw[normal_edge] (\x,\y) -- (\x+1,\y);
    }
}

% Draw all vertical edges in gray
\foreach \x in {0,1,2,3,4} {
    \foreach \y in {0,1,2,3,4} {
        \draw[normal_edge] (\x,\y) -- (\x,\y+1);
    }
}

\draw[red_edge] (3,3) -- (3,4);
\draw[red_edge] (4,3) -- (4,4);
\draw[blue_edge] (4,4) -- (3,4);
\draw[blue_edge] (3,4) -- (2,4);
\draw[blue_edge] (2,4) -- (2,3);

\draw[red_edge] (3,2) -- (3,1);   
\draw[blue_edge] (3,1) -- (2,1);  
\draw[red_edge] (4,2) -- (4,1);   
\draw[blue_edge] (4,1) -- (3,1);

\draw[blue, line width=3pt] (2.15,1.15) rectangle (2.85,1.85);
\draw[blue, line width=3pt] (3.15,1.15) rectangle (3.85,1.85);
% Draw all vertices
\foreach \x in {0,1,2,3,4} {
    \foreach \y in {0,1,2,3,4,5} {
        \node[vertex] at (\x,\y) {};
    }
}
\end{tikzpicture}
\caption{Two representations of a string automorphism $m\in M\cap N$ by a circuit supported on the right half-line. The bottom one is manifestly gauge-invariant.}
\end{figure}
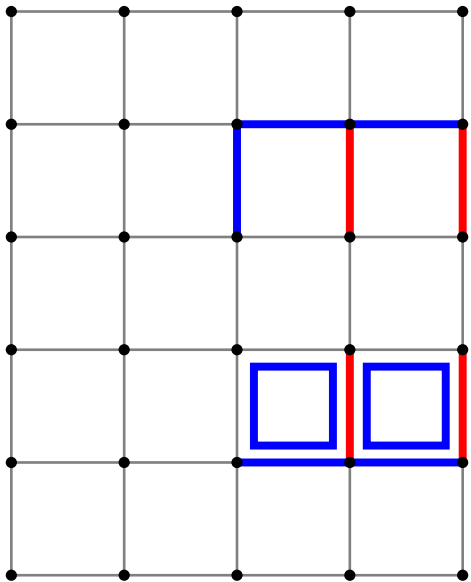

The equivalence class of $\mathbf{x}=(m,m^{-1})\in Q$ defines a nontrivial element of $\pi_2$, because it flips the sign of a single $B_p$ (as well as some $U_e$), and conjugation with a local gauge-invariant observable cannot do this. Thus we defined a distinguished $\ZZ_2$ 1-form symmetry of the twisted $\ZZ_2$ gauge theory.

We can compute the anomaly of this 1-form symmetry by evaluating $\{\mathbf{x},\mathbf{x}\}$, or equivalently the "commutator" of the circuits shown in Fig. 6 and Fig. 7. It is easy to see that the "commutator" is $-1$, thus the anomaly is the quadratic function $q(\mathbf{x}^n)=\exp(i\pi n^2)$, where $\mathbf{x}$ is the generator of $\ZZ_2$. 
%Here we identify $U(1)$ with $\RR/\ZZ$. 

One can think of this quadratic function as describing a nontrivial ribbon category of topological defect lines implementing the $\ZZ_2$ 1-form symmetry. This ribbon category is equivalent to the category of finite-dimensional super-vector spaces, with the generator of $\ZZ_2$ symmetry corresponding to an odd one-dimensional vector space. This is what makes the twisted $\ZZ_2$ gauge theory dual to a theory of fermions.

There exists a 3d Euclidean $\ZZ_2$ gauge theory with the same anomaly. On a triangulated 3-manifold, its action is 
\begin{align}\label{eq:actionZ2}
    S=\frac{1}{g^2}\sum_p (\delta a)_p +i\pi \int a\cup\delta a,
\end{align}
Here $a$ is a 1-cochain valued in $\ZZ_2=\{0,1\}$. This theory has a global 1-form symmetry $a\mapsto a+\alpha$, where $\alpha$ is a 1-cocycle. This symmetry is anomalous, with the anomaly described by a 4d topological action
\begin{align}
    S_{top}=i\pi \int B\cup B,
\end{align}
where $B$ is 2-form gauge field valued in $\ZZ_2$ (i.e. a 2-cocycle with values in $\ZZ_2$). We conjecture that in the continuum limit, this Euclidean theory is equivalent to the Hamiltonian gauge theory discussed above. The precise comparison is subtle, since the second term in the action (\ref{eq:actionZ2}) assumes a triangulated 3-manifold, while the Hamiltonian approach assumes that the space-time is a product of space and time. 

Note that the bilinear form $b(x,y)=q(xy)/q(x)q(y)$ is trivial (identically $1$). Thus, the category of topological defect lines is not modular. In the Euclidean theory, the expectation value of the Hopf link of two defect lines is $1$ and cannot be used to detect the anomaly. Similarly, the commutator of the circuit shown in Fig. 5 and a similar circuit in the vertical direction is trivial and also cannot be used to detect the anomaly. 

\subsection{Twisted $\ZZ_n$ gauge theory}

It is straightforward to generalize the discussion to a $\ZZ_n$ gauge theory. The corresponding Hamiltonian lattice model on a square lattice has a "qunit" (a Hilbert space of dimension $n$) on each edge. The corresponding algebra of observables is generated by unitary clock and shift operators $T,S$, which satisfy the relations
\begin{align}
    T^n=S^n=1,\quad ST=\lambda TS,
\end{align}
where $\lambda=\exp(2\pi i/n)$. There is one such pair of generators for each oriented edge $e$; reversing orientation replaces $S,T$ with their inverses. We identify an eigenstate of $T_e$ with a state where the vector potential $a(e)$ has a definite value. The modified Gauss law constraints are 
\begin{align}
    B^s_{NE(v)}\prod_{e\supset v} S_e=1,
\end{align}
where $B_p=\prod_{e\subset p} T_e$, and our orientation conventions are shown in Fig. 8. 

\begin{figure}
\centering
\begin{tikzpicture}[   
    scale=1.5,
    normal_edge/.style={gray, line width=1pt},
    red_edge/.style={red, line width=3pt},
    blue_edge/.style={blue, line width=3pt},
    vertex/.style={circle, fill=black, inner sep=1.5pt}
]
\tikzset{
  midarrow/.style={
    decoration={markings, mark=at position 0.7 with {\arrow{stealth}}},
    postaction={decorate}
  }
}

% Draw all horizontal edges in gray
\foreach \x in {0,1,2,3} {
    \foreach \y in {0,1,2,3,4} {
        \draw[normal_edge] (\x,\y) -- (\x+1,\y);
    }
}

% Draw all vertical edges in gray
\foreach \x in {0,1,2,3,4} {
    \foreach \y in {0,1,2,3} {
        \draw[normal_edge] (\x,\y) -- (\x,\y+1);
    }
}

% Draw the four oriented red edges that share a vertex (choosing vertex at (2,2))
% These will overdraw the gray edges
\draw[midarrow,red_edge] (2,2) -- (2,3);   % up
\draw[midarrow,red_edge] (2,2) -- (2,1);   % down
\draw[midarrow,red_edge] (2,2) -- (1,2);   % left
\draw[midarrow,red_edge] (2,2) -- (3,2);   % right

%Draw a blue plaquette
\draw[midarrow,blue_edge] (2.2,2.2) -- (2.8,2.2);   % up
\draw[midarrow,blue_edge] (2.8,2.2) -- (2.8,2.8);   % down
\draw[midarrow,blue_edge] (2.8,2.8) -- (2.2,2.8);   % left
\draw[midarrow,blue_edge] (2.2,2.8) -- (2.2,2.2);   % right

\node[] at (2.5,2.5) {$\color{blue} s$};

% Draw all vertices
\foreach \x in {0,1,2,3,4} {
    \foreach \y in {0,1,2,3,4} {
        \node[vertex] at (\x,\y) {};
    }
}

% Add blue square centered at the northeast lattice square
% The northeast square from vertex (2,2) has center at (2.5,2.5)
% Make it slightly smaller than the unit square
%\draw[blue, line width=3pt] (2.15,2.15) rectangle (2.85,2.85);

\end{tikzpicture}

\caption{A graphic representation of the constraint in twisted $\ZZ_n$ gauge theory. An edge marked in red (resp. blue) denotes $S$ (resp. $T$). The blue square with $s$ inside denotes $B_{NE(v)}$ raised to the power $s$.}
\end{figure}
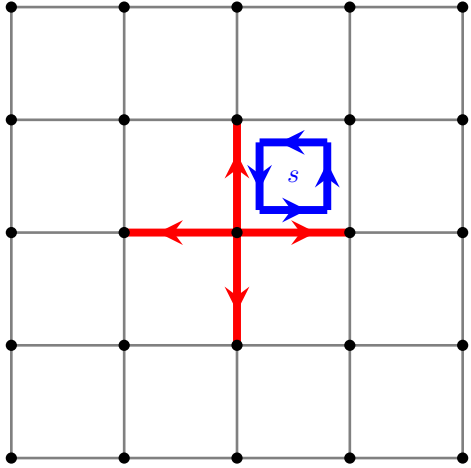

A gauge-invariant circuit acting trivially on gauge-invariant observables is shown in Fig. 9, and its gauge-invariant truncations to the left and right half-lines in Fig. 10 and Fig. 11. They define a distinguished $\ZZ_n$ subgroup of $\pi_2$. The "commutator" of the truncations is $\lambda^{-s}$. Thus, the quadratic function $q$ is
\begin{align}\label{eq:quadZn}
q(\mathbf{x}^m)=\exp\left(\frac{2 i \pi s m^2}{n}\right),
\end{align}
where $\mathbf{x}$ is the generator of $\ZZ_n$.
\begin{figure}
\centering
\begin{tikzpicture}[
    scale=1.5,
    normal_edge/.style={gray, line width=1pt},
    red_edge/.style={red, line width=3pt},
    blue_edge/.style={blue, line width=3pt},
    vertex/.style={circle, fill=black, inner sep=1.5pt}
]
\tikzset{
  midarrow/.style={
    decoration={markings, mark=at position 0.7 with {\arrow{stealth}}},
    postaction={decorate}
  }
}

% Draw all horizontal edges in gray
\foreach \x in {0,1,2,3} {
    \foreach \y in {0,1,2,3} {
        \draw[normal_edge] (\x,\y) -- (\x+1,\y);
    }
}

% Draw all vertical edges in gray
\foreach \x in {0,1,2,3,4} {
    \foreach \y in {0,1,2} {
        \draw[normal_edge] (\x,\y) -- (\x,\y+1);
    }
}

\draw[midarrow,red_edge] (0,1) -- (0,2); 
\draw[midarrow,red_edge] (1,1) -- (1,2);  
\draw[midarrow,red_edge] (2,1) -- (2,2);
\draw[midarrow,red_edge] (3,1) -- (3,2); 
\draw[midarrow,red_edge] (4,1) -- (4,2); 

\draw[midarrow,blue_edge] (1,2) -- (0,2);
\node[] at (0.5,2.2) {$\color{blue} s$};
\draw[midarrow,blue_edge] (2,2) -- (1,2);
\node[] at (1.5,2.2) {$\color{blue} s$};
\draw[midarrow,blue_edge] (3,2) -- (2,2);
\node[] at (2.5,2.2) {$\color{blue} s$};
\draw[midarrow,blue_edge] (4,2) -- (3,2);
\node[] at (3.5,2.2) {$\color{blue} s$};

% Draw all vertices
\foreach \x in {0,1,2,3,4} {
    \foreach \y in {0,1,2,3} {
        \node[vertex] at (\x,\y) {};
    }
}
\end{tikzpicture}
\caption{A gauge-invariant circuit in the twisted $\ZZ_n$ theory which acts by the identity automorphism on gauge-invariant observables.}
\end{figure}
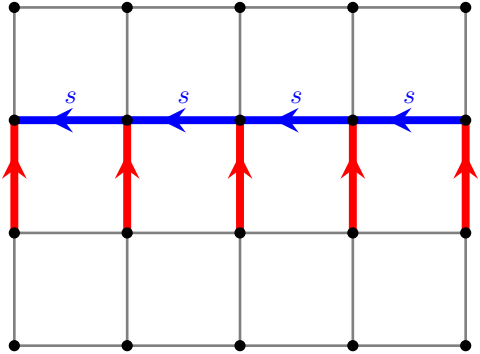

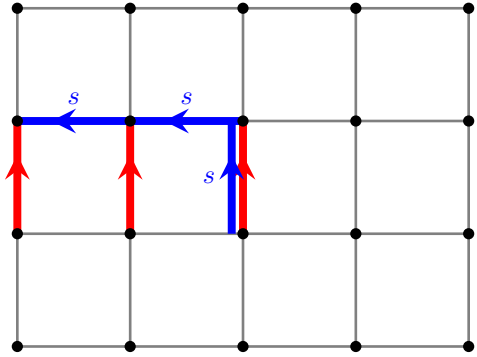
\begin{figure}
\centering
\begin{tikzpicture}[
    scale=1.5,
    normal_edge/.style={gray, line width=1pt},
    red_edge/.style={red, line width=3pt},
    blue_edge/.style={blue, line width=3pt},
    vertex/.style={circle, fill=black, inner sep=1.5pt}
]
\tikzset{
  midarrow/.style={
    decoration={markings, mark=at position 0.7 with {\arrow{stealth}}},
    postaction={decorate}
  }
}

% Draw all horizontal edges in gray
\foreach \x in {0,1,2,3} {
    \foreach \y in {0,1,2,3} {
        \draw[normal_edge] (\x,\y) -- (\x+1,\y);
    }
}

% Draw all vertical edges in gray
\foreach \x in {0,1,2,3,4} {
    \foreach \y in {0,1,2} {
        \draw[normal_edge] (\x,\y) -- (\x,\y+1);
    }
}

\draw[midarrow,red_edge] (0,1) -- (0,2); 
\draw[midarrow,red_edge] (1,1) -- (1,2);  
\draw[midarrow,red_edge] (2,1) -- (2,2); 

\draw[midarrow,blue_edge] (1,2) -- (0,2);
\node[] at (0.5,2.2) {$\color{blue} s$};
\draw[midarrow,blue_edge] (2,2) -- (1,2);
\node[] at (1.5,2.2) {$\color{blue} s$};
\draw[midarrow,blue_edge] (1.9,1) -- (1.9,2);
\node[] at (1.7,1.5) {$\color{blue} s$};

% Draw all vertices
\foreach \x in {0,1,2,3,4} {
    \foreach \y in {0,1,2,3} {
        \node[vertex] at (\x,\y) {};
    }
}
\end{tikzpicture}
\caption{A representations of the string automorphism $m\in M\cap N$ by a gauge-invariant circuit supported on the left half-line.}
\end{figure}

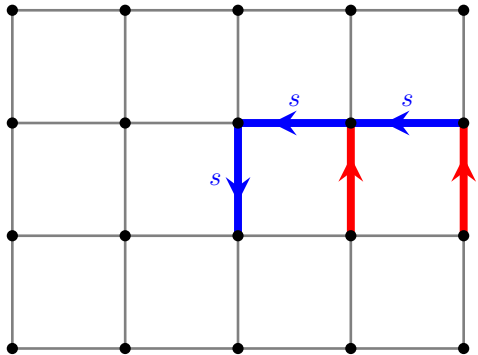
\begin{figure}
\centering
\begin{tikzpicture}[
    scale=1.5,
    normal_edge/.style={gray, line width=1pt},
    red_edge/.style={red, line width=3pt},
    blue_edge/.style={blue, line width=3pt},
    vertex/.style={circle, fill=black, inner sep=1.5pt}
]
\tikzset{
  midarrow/.style={
    decoration={markings, mark=at position 0.7 with {\arrow{stealth}}},
    postaction={decorate}
  }
}

% Draw all horizontal edges in gray
\foreach \x in {0,1,2,3} {
    \foreach \y in {0,1,2,3} {
        \draw[normal_edge] (\x,\y) -- (\x+1,\y);
    }
}

% Draw all vertical edges in gray
\foreach \x in {0,1,2,3,4} {
    \foreach \y in {0,1,2} {
        \draw[normal_edge] (\x,\y) -- (\x,\y+1);
    }
}

\draw[midarrow,red_edge] (3,1) -- (3,2); 
\draw[midarrow,red_edge] (4,1) -- (4,2);

\draw[midarrow,blue_edge] (3,2) -- (2,2);
\node[] at (2.5,2.2) {$\color{blue} s$};
\draw[midarrow,blue_edge] (4,2) -- (3,2);
\node[] at (3.5,2.2) {$\color{blue} s$};
\draw[midarrow,blue_edge] (2,2) -- (2,1);
\node[] at (1.8,1.5) {$\color{blue} s$};

% Draw all vertices
\foreach \x in {0,1,2,3,4} {
    \foreach \y in {0,1,2,3} {
        \node[vertex] at (\x,\y) {};
    }
}
\end{tikzpicture}
\caption{A representations of the string automorphism $m\in M\cap N$ by a gauge-invariant circuit supported on the right half-line.}
\end{figure}

A Euclidean 3d $\ZZ_n$ gauge theory with the same anomaly has an action
\begin{align}
    S=\frac{1}{g^2}\sum_p (\delta a)_p +\frac{2\pi is}{n} \int a\cup\delta a.
\end{align}
Here $a$ is a $\ZZ_n$-valued 1-cochain on a triangulated 3-manifold and $s\in \ZZ_n$ is a parameter. Repeating the computation of the anomaly action, one finds that it is given by
\begin{align}
    S_{top}=\frac{2\pi is}{n}\int B\cup B,
\end{align}
where $B$ is a 2-cocycle with values in $\ZZ_n$. The corresponding quadratic function is (\ref{eq:quadZn}). We conjecture that in the continuum limit the Euclidean and Hamiltonian theories are equivalent.

The bilinear form on $\ZZ_n$ corresponding to the above quadratic function is $b(\mathbf{x}^m,\mathbf{x}^k)=\exp(4i\pi smk/n)$. It can be detected in the Euclidean approach by computing the expectation value of the Hopf link of two topological defect lines, or in the operator approach by computing the commutator of two perpendicular infinite circuits as in Fig. 9. For odd $n$ the bilinear form uniquely determines the quadratic function, but for even $n$ this is no longer true, as we saw in the previous subsection. The prescription for computing $q$ given by the crossed square in effect provides an unambiguous value for the square root of the commutator of two infinite circuits.

\subsection{$\ZZ_n\times \ZZ_n$ gauge theory}

In this section we examine a $\ZZ_n\times\ZZ_n$ 2+1d lattice gauge theory with a modified Gauss law. In the Hamiltonian approach, there is a pair of qunits on each edge. We denote the corresponding clock and shift generators $T_e,S_e$ and $\hT_e,\hS_e$.  The Gauss law constraints are:
\begin{align}
    \hB^s_{NE(v)}\prod_{e\supset v} S_e=1 , \quad B^s_{NE(v)}\prod_{e\supset v} \hS_e=1.
\end{align}
Here $\hB_p=\prod_{e\subset p}\hT_e$ and $s\in \ZZ_n$ is a parameter. A graphic representation of the Gauss laws is shown in Fig. 12. 

\begin{figure}
\centering
\begin{tikzpicture}[
    scale=1.5,
    normal_edge/.style={gray, line width=1pt},
    red_edge/.style={red, line width=3pt},
    dashed_red_edge/.style={red, dashed, line width=3pt},
    vertex/.style={circle, fill=black, inner sep=1.5pt},
    blue_edge/.style={blue, line width=3pt},
    dashed_blue_edge/.style={blue, dashed, line width=3pt},
]
\tikzset{
  midarrow/.style={
    decoration={markings, mark=at position 0.7 with {\arrow{stealth}}},
    postaction={decorate}
  }
}

% Draw all horizontal edges in gray
\foreach \x in {0,1,2,3,4,5,6} {
    \foreach \y in {0,1,2,3,4} {
        \draw[normal_edge] (\x,\y) -- (\x+1,\y);
    }
}

% Draw all vertical edges in gray
\foreach \x in {0,1,2,3,4,5,6,7} {
    \foreach \y in {0,1,2,3} {
        \draw[normal_edge] (\x,\y) -- (\x,\y+1);
    }
}

\draw[midarrow,red_edge] (2,2) -- (2,3);   % up
\draw[midarrow,red_edge] (2,2) -- (2,1);   % down
\draw[midarrow,red_edge] (2,2) -- (1,2);   % left
\draw[midarrow,red_edge] (2,2) -- (3,2);   % right

%Draw a dashed blue plaquette
\draw[midarrow,dashed_blue_edge] (2.2,2.2) -- (2.8,2.2);   % up
\draw[midarrow,dashed_blue_edge] (2.8,2.2) -- (2.8,2.8);   % down
\draw[midarrow,dashed_blue_edge] (2.8,2.8) -- (2.2,2.8);   % left
\draw[midarrow,dashed_blue_edge] (2.2,2.8) -- (2.2,2.2);   % right

\node[] at (2.5,2.5) {$\color{blue} s$};

% Draw the four red edges that share a vertex (choosing vertex at (5,2))
% These will overdraw the gray edges
\draw[midarrow,dashed_red_edge] (5,2) -- (5,3);   % up
\draw[midarrow,dashed_red_edge] (5,2) -- (5,1);   % down
\draw[midarrow,dashed_red_edge] (5,2) -- (4,2);   % left
\draw[midarrow,dashed_red_edge] (5,2) -- (6,2);   % right

%Draw a blue plaquette
\draw[midarrow,blue_edge] (5.2,2.2) -- (5.8,2.2);   % up
\draw[midarrow,blue_edge] (5.8,2.2) -- (5.8,2.8);   % down
\draw[midarrow,blue_edge] (5.8,2.8) -- (5.2,2.8);   % left
\draw[midarrow,blue_edge] (5.2,2.8) -- (5.2,2.2);   % right

\node[] at (5.5,2.5) {$\color{blue} s$};

% Draw all vertices
\foreach \x in {0,1,2,3,4,5,6,7} {
    \foreach \y in {0,1,2,3,4} {
        \node[vertex] at (\x,\y) {};
    }
}

\end{tikzpicture}

\caption{A graphic representation of the Gauss law constraints in the $\ZZ_n\times\ZZ_n$ gauge theory. Dashed (resp. solid) lines correspond to generators with hats (resp. without hats). Red edges correspond to $S$ generators, blue edges correspond to $T$ generators.}
\end{figure}
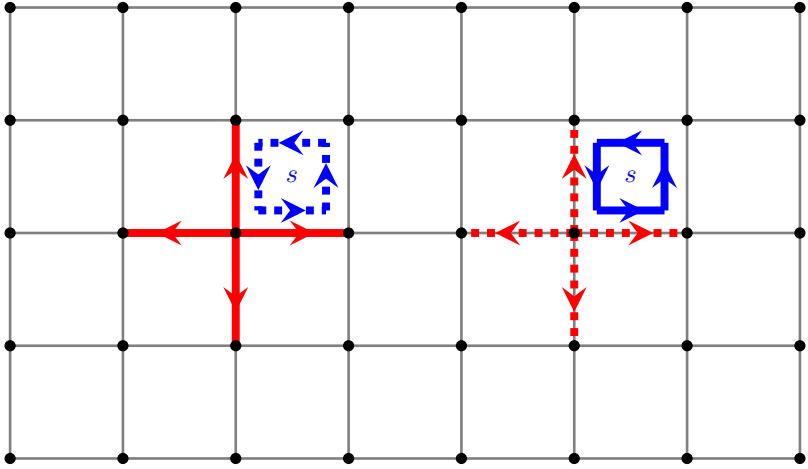

The two Gauss laws give rise to two gauge-invariant circuits that act by identity automorphisms on all gauge-invariant local observables. They are depicted in Fig. 13. 

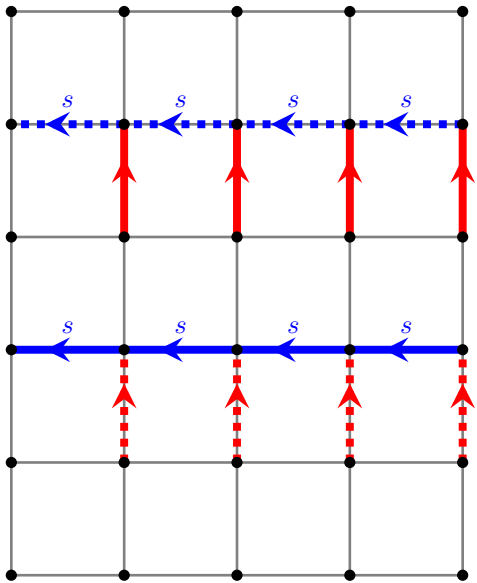
\begin{figure}
\centering
\begin{tikzpicture}[
    scale=1.5,
    normal_edge/.style={gray, line width=1pt},
    red_edge/.style={red, line width=3pt},
    dashed_red_edge/.style={red, dashed, line width=3pt},
    vertex/.style={circle, fill=black, inner sep=1.5pt},
    blue_edge/.style={blue, line width=3pt},
    dashed_blue_edge/.style={blue, dashed, line width=3pt},
]
\tikzset{
  midarrow/.style={
    decoration={markings, mark=at position 0.7 with {\arrow{stealth}}},
    postaction={decorate}
  }
}
% Draw all horizontal edges in gray
\foreach \x in {0,1,2,3} {
    \foreach \y in {0,1,2,3,4,5} {
        \draw[normal_edge] (\x,\y) -- (\x+1,\y);
    }
}

% Draw all vertical edges in gray
\foreach \x in {0,1,2,3,4} {
    \foreach \y in {0,1,2,3,4} {
        \draw[normal_edge] (\x,\y) -- (\x,\y+1);
    }
}

\foreach \x in {1,2,3,4} {
    \draw[midarrow,dashed_red_edge] (\x,1) -- (\x,2); 
    \draw[midarrow,red_edge] (\x,3) -- (\x,4);
}

\foreach \x in {1,2,3,4} {
    \draw[midarrow,blue_edge] (\x,2) -- (\x-1,2);
    \draw[midarrow,dashed_blue_edge] (\x,4) -- (\x-1,4);
    \node[] at (\x-0.5,2.2) {$\color{blue} s$};
    \node[] at (\x-0.5,4.2) {$\color{blue} s$};
}

% Draw all vertices
\foreach \x in {0,1,2,3,4} {
    \foreach \y in {0,1,2,3,4,5} {
        \node[vertex] at (\x,\y) {};
    }
}
\end{tikzpicture}
\caption{Two circuits in $\ZZ_n\times\ZZ_n$ theory which act by the identity automorphism on gauge-invariant observables.}
\end{figure}

Truncating the two circuits in a gauge-invariant way, we obtain the string automorphisms shown in Fig. 14 and Fig. 15. These string automorphisms generate a $\ZZ_n\times\ZZ_n$ subgroup of $\pi_2$. Thus the theory has a $\ZZ_n\times\ZZ_n$ 1-form symmetry. Repeating the manipulations of the previous subsection and computing the commutators of all three pairs of left and right-localized circuits, we obtain the quadratic function on $\ZZ_n\times\ZZ_n$ which encodes the anomaly:
\begin{align}\label{eq:ZnZn}
    q(\mathbf{x}^m \hat{\mathbf{x}}^k)=\exp\left(\frac{4i\pi s mk}{n}\right),
\end{align}
where $\mathbf{x}, \hat{\mathbf{x}}$ are generators of $\ZZ_n$ factors in $\ZZ_n\times\ZZ_n$. 

For $n=2$ it vanishes, but in general it is nonzero. If $n$ is odd and $s$ does not divide $n,$ the corresponding bilinear form is non-degenerate and corresponds to a modular tensor category of topological defect lines.

\begin{figure}
\centering
\begin{tikzpicture}[
    scale=1.5,
    normal_edge/.style={gray, line width=1pt},
    red_edge/.style={red, line width=3pt},
    dashed_red_edge/.style={red, dashed, line width=3pt},
    vertex/.style={circle, fill=black, inner sep=1.5pt},
    blue_edge/.style={blue, line width=3pt},
    dashed_blue_edge/.style={blue, dashed, line width=3pt},
]
\tikzset{
  midarrow/.style={
    decoration={markings, mark=at position 0.7 with {\arrow{stealth}}},
    postaction={decorate}
  }
}
% Draw all horizontal edges in gray
\foreach \x in {0,1,2,3} {
    \foreach \y in {0,1,2,3,4,5} {
        \draw[normal_edge] (\x,\y) -- (\x+1,\y);
    }
}

% Draw all vertical edges in gray
\foreach \x in {0,1,2,3,4} {
    \foreach \y in {0,1,2,3,4} {
        \draw[normal_edge] (\x,\y) -- (\x,\y+1);
    }
}

\draw[midarrow,dashed_red_edge] (0,1) -- (0,2); 
\draw[midarrow,dashed_red_edge] (1,1) -- (1,2);  
\draw[midarrow,dashed_red_edge] (2,1) -- (2,2);

\draw[midarrow,blue_edge] (1,2) -- (0,2);
\draw[midarrow,blue_edge] (2,2) -- (1,2);
\draw[midarrow,blue_edge] (1.9,1) -- (1.9,2);

\node[] at (1.7,1.5) {$\color{blue} s$};
\node[] at (1.5,2.2) {$\color{blue} s$};
\node[] at (0.5,2.2) {$\color{blue} s$};

\draw[midarrow,red_edge] (0,3) -- (0,4); 
\draw[midarrow,red_edge] (1,3) -- (1,4);   
\draw[midarrow,red_edge] (2,3) -- (2,4);

\draw[midarrow,dashed_blue_edge] (1,4) -- (0,4);
\draw[midarrow,dashed_blue_edge] (2,4) -- (1,4);
\draw[midarrow,dashed_blue_edge] (1.9,3) -- (1.9,4);

\node[] at (1.7,3.5) {$\color{blue} s$};
\node[] at (1.5,4.2) {$\color{blue} s$};
\node[] at (0.5,4.2) {$\color{blue} s$};

% Draw all vertices
\foreach \x in {0,1,2,3,4} {
    \foreach \y in {0,1,2,3,4,5} {
        \node[vertex] at (\x,\y) {};
    }
}
\end{tikzpicture}
\caption{A representation of the string automorphisms in $\ZZ_n\times\ZZ_n$ theory by circuits supported on the left half-line.}
\end{figure}

\begin{figure}
\centering
\begin{tikzpicture}[
    scale=1.5,
    normal_edge/.style={gray, line width=1pt},
    red_edge/.style={red, line width=3pt},
    dashed_red_edge/.style={red, dashed, line width=3pt},
    vertex/.style={circle, fill=black, inner sep=1.5pt},
    blue_edge/.style={blue, line width=3pt},
    dashed_blue_edge/.style={blue, dashed, line width=3pt},
]
\tikzset{
  midarrow/.style={
    decoration={markings, mark=at position 0.7 with {\arrow{stealth}}},
    postaction={decorate}
  }
}
% Draw all horizontal edges in gray
\foreach \x in {0,1,2,3} {
    \foreach \y in {0,1,2,3,4,5} {
        \draw[normal_edge] (\x,\y) -- (\x+1,\y);
    }
}

% Draw all vertical edges in gray
\foreach \x in {0,1,2,3,4} {
    \foreach \y in {0,1,2,3,4} {
        \draw[normal_edge] (\x,\y) -- (\x,\y+1);
    }
}

\draw[midarrow,dashed_red_edge] (3,1) -- (3,2);
\draw[midarrow,dashed_red_edge] (4,1) -- (4,2);

\node[] at (1.8,1.5) {$\color{blue} s$};
\node[] at (2.5,2.2) {$\color{blue} s$};
\node[] at (3.5,2.2) {$\color{blue} s$};

\draw[midarrow,blue_edge] (3,2) -- (2,2);
\draw[midarrow,blue_edge] (4,2) -- (3,2);
\draw[midarrow,blue_edge] (2,2) -- (2,1);

\draw[midarrow,red_edge] (3,3) -- (3,4);   
\draw[midarrow,red_edge] (4,3) -- (4,4);  

\node[] at (1.8,3.5) {$\color{blue} s$};
\node[] at (2.5,4.2) {$\color{blue} s$};
\node[] at (3.5,4.2) {$\color{blue} s$};

\draw[midarrow,dashed_blue_edge] (3,4) -- (2,4);
\draw[midarrow,dashed_blue_edge] (4,4) -- (3,4);
\draw[midarrow,dashed_blue_edge] (2,4) -- (2,3);

% Draw all vertices
\foreach \x in {0,1,2,3,4} {
    \foreach \y in {0,1,2,3,4,5} {
        \node[vertex] at (\x,\y) {};
    }
}
\end{tikzpicture}
\caption{A representation of the string automorphisms in $\ZZ_n\times\ZZ_n$ theory by circuits supported on the right half-line.}
\end{figure}

The $\ZZ_n\times\ZZ_n$ gauge theory has a $\ZZ_2$ symmetry which exchanges the two $\ZZ_n$ gauge fields. A Euclidean counterpart with this property would have an action
\begin{align}
    S=\frac{1}{g^2}\sum_p \left((\delta a)_p+(\delta \ha)_p\right) +\frac{2\pi is}{n} \int (a\cup\delta \ha+\ha\cup\delta a),
\end{align}
where $a,\ha$ are $\ZZ_n$-valued 1-cochains. It has a 1-form symmetry $\ZZ_n\times\ZZ_n$, corresponding to the invariance of the action under addition of 1-cocycles to $a$ and $\ha$. It is also manifestly invariant under the exchange of $a$ and $\ha$.\footnote{Note that the cup product is not graded symmetric on the level cochains, even though it induces a graded symmetric product of cohomology classes. Thus $\int a\cup\delta \ha+\int \ha\cup\delta a$ is not equal to $2\int a\cup\delta \ha$, and the latter expression is not invariant under the exchange of $a$ and $\ha$.} The anomaly action is
\begin{align}
    \frac{2\pi is}{n}\int (B\cup \hB+\hB\cup B)=\frac{4\pi is}{n}\int B\cup \hB,
\end{align}
where $B,\hB$ are 2-cocycles with values in $\ZZ_n$. The corresponding quadratic function on $\ZZ_n\times\ZZ_n$ is (\ref{eq:ZnZn}). We conjecture that this Euclidean theory is equivalent to the above Hamiltonian theory in the continuum limit.

\section{Generalization to higher dimensions}

This section proposes a generalization of the preceding analysis to an arbitrary number of spatial dimensions. The discussion is a sketch that requires more detailed development in future work.

A general finite homotopy type can be algebraically encoded in a structure known as a crossed $n$-cube \cite{crossedcubes}. Denote a set of integers from 1 to $n$ as $\langle n \rangle$. A crossed $n$-cube is a family of groups $M_A$ (where $A\subset \langle n \rangle $) together with the morphisms $\mu_i: M_A \rightarrow M_{A\backslash \{i\}}$ and functions $h: M_A \times M_B \rightarrow M_{A \bigcup B}$ satisfying conditions listed below. If $A \subset B$, denote by ${}^a b = h(a,b)b$ for $a \in M_A$ and $b \in M_B$. For $a,a' \in M_A$, $b,b' \in M_B$, $c \in M_c$ and $i,j \in \langle n \rangle $, the morphisms $\mu$ and functions $h$ satisfy
\begin{align}
    &\mu_i a = a  \text{ if } i \notin A,\\
    &\mu_i \mu_j a =\mu_j \mu_i a ,\\
    &\mu_ih(a,b) = h(\mu_ia,\mu_ib),\\
    &h(a,b) = h(a,\mu_i b)= h(\mu_i a,b) \text{ if } i \in A \text{ and } i \in B,\\
    &h(a,a')=[a,a'],\\
    &h(a,b)= h(b,a)^{-1},\\
    &h(a,b) = 1 \text{ if } a=1 \text{ or } b = 1,\\
    &h(aa',b) ={}^ah(a',b)h(a,b),\\
    &h(a,b b') =h(a,b){}^b h(a,b'),\\
    &{}^a h(h(a^{-1},b),c){}^c h(h(c^{-1},a),b){}^b h(h(b^{-1},c),a)=1,\\
    &{}^a h(b,c)=h({}^a b,{}^a c) \text{ if } A\subset B \text{ and } A\subset C. 
\end{align}

A simple example of a crossed $n$-cube is a family of normal subgroups $M_{\{1\}},\dots,M_{\{n\}}$ of a group $M_{\emptyset}$ with $M_A$ being the intersection of subgroups indexed by $A$. The morphisms are inclusions, and the functions $h$ are given by commutators. Note that a crossed 1-cube is the same as a crossed module, while a crossed 2-cube is the same as a crossed square.

A crossed $n$-cube naturally appears when one considers automorphisms approximately localized on spatial subsets. Take a cover of $d$-dimensional space $\mathbb R^d$ by $n$ (closed) sets $U_i$. Define $M_A=\KT(\bigcap_{i \in A} U_i)$ as the group of approximately localized automorphisms at the intersection $\bigcap_{i \in A} U_i$ if the intersection is unbounded and as the group of local unitary operators if it is bounded. The $M_{\emptyset}$ is the group of all automorphisms. The morphisms $\mu_i$ are $ \ u\rightarrow \Ad_u$ if they map local operators into automorphisms and inclusions otherwise. Similarly, the functions $h$ are defined as commutators or "commutators", whichever is appropriate.   

This crossed $n$-cube encodes obstructions for representations of automorphism as a product of automorphisms approximately localized on subregions. There are infinitely many choices of cover, but by refining it one can reduce the problem to a canonical choice. 

First, we don't want the topology of the sets $U_i$ themselves to complicate the picture. Therefore, we restrict to the case where $U_i$ are infinite cones and the bases of the cones, as well as any of their intersections, are contractible at infinity. Second, by taking a suitable refinement, we can assume that cones intersect only along their boundaries.  Lastly, by subdividing the regions, we get a triangulation of the $(d-1)$-sphere at infinity. (A minimal canonical way to construct such a cover is to consider a $d$-simplex and define $d$ sets $U_i$ as cones from its center to each of the $(d-1)$-dimensional faces.)

This yields a crossed $n$-cube, where $n$ can be pretty large. However, the non-trivial information is contained in smaller $d$-cubes: for each simplex in the triangulation, we can take $\widetilde M_\emptyset$ to be the automorphisms localized on it and $\widetilde M_{\{i\}}$ for $i=1,\dots, d$ automorphisms localized on its ($d-1$)-dimensional faces. Whenever one has a homomorphism from a group $G$ to automorphisms acting on the whole space, one can restrict\footnote{There is an obstruction for this restriction taking values in the group of $M_\emptyset/(M_{\{1\}}\rtimes\dots \rtimes M_{\{n\}})$. In the argument, we assume that it is trivial.  } it to a homomorphism from $G$ to the factor group $\widetilde M_\emptyset/(\widetilde M_{\{1\}}\rtimes\dots \rtimes\widetilde M_{\{n\}})$ of automorphisms localized on the simplex modulo automorphisms localized along its boundary.

From this crossed $d$-cube, one can construct (see section 3 of \cite{LODAY}) a complex of length $d+1$:
\begin{equation}
    C_d \xrightarrow{\partial_d} C_{d-1}\xrightarrow{\partial_{d-1}}\dots  \xrightarrow{\partial_1} C_0,
\end{equation}
where $C_k$ is a semidirect product of $\widetilde M_A$ with $|A|=k$ and $d_k$ are combinations of $\widetilde\mu_i$. The images ${\rm im}\, \partial_k$  are normal subgroups of $C_{k-1}$ and one can define the groups $\pi_1(C_*)={{\rm coker}\, \partial_{1}}$, $\pi_k(C_*)={{\rm ker}\, \partial_{k-1}}/{{\rm im}\, \partial_{k}}$ for $k=2,\dots,d$ and $\pi_d(C_*)={{\rm ker}\, \partial_{d}}$. 

As in the two-dimensional case, $\pi_k(C_*)$ can be identified with higher-form symmetries. The functions $h(a,b)$ induce maps between the homotopy group and correspond to non-trivial higher-group structure. It can be used to detect anomalies.

\section{Concluding remarks}

In this paper, we proposed a rigorous definition of higher-form symmetries of quantum gauge theories on a spatial lattice. We saw that the origin of higher symmetry is the extra structure present in any many-body system: the notion of a symmetry localized in a spatial region. Similar considerations should apply to continuum QFT in the Haag-Kastler approach \cite{HaagKastler}. Thus, incorporating higher-form symmetry does not seem to require a conceptually new axiomatization of QFT. We also hope that our approach will pave the way to a proof of some conjectures, such as the higher-form generalization of the Hohenberg-Mermin-Wagner theorem \cite{gensym}.

From a mathematical viewpoint, the notion of a symmetry localized in a region can be encoded in a precosheaf of groups with some additional structure (the "commutator" map). This precosheaf attaches to a region $U$ the group $\KT(U)$ of symmetries localized in $U$; for every inclusion of regions $U
\subseteq V$ one has a group homomorphism $\KT(U)\ra \KT(V)$ which sends $\KT(U)$ to a normal subgroup of $\KT(V)$. Upon choosing a finite cover of the space, a sort of nonabelian \Cech\ construction gives a crossed $n$-cube where $n$ depends on the number of sets in the cover. Presumably, the nonabelian \Cech\ construction is functorial in a suitable sense, ensuring that the precise choice of the cover does not matter if it is sufficiently nice. 

All constructions in this paper are algebraic, so the topology of the symmetry group does not play a role. Equivalently, all symmetry groups have discrete topology. For physics applications, it is important to consider higher-form Lie group symmetry. 
We note that on the infinitesimal level, symmetries localizable in regions form a precosheaf of Lie algebras with additional structure. It was shown in \cite{AKY} how to attach a differential graded Lie algebra to such a precosheaf and a choice of a cover. A differential graded Lie algebra over rational numbers is an "infinitesimal" (rational) version of a homotopy type, as first shown by Quillen \cite{Quillen}. It is very suggestive that the same type of mathematical structure arises by applying a nonabelian \Cech\ construction to infinitesimal symmetries of quantum lattice systems.

\appendix
\renewcommand{\thesection}{}
\makeatletter
\renewcommand{\@seccntformat}[1]{}
\makeatother
\section{Appendix}

In this Appendix we will show that for $\partial x =\partial y =\partial z = 1$ the functions $q(x)$ and  $b(x,y)=q(xy)q(x)^{-1}q(y)^{-1}$ satisfy
\begin{align}
    & \delta q(x) = 1,\\
    &q(x \delta l)=q(\delta l x)=q(x),\\
    &q(x^n) = q(x)^{n^2},\\
    &b(x,yz)= b(x,y)b(x,z),\\
    &b(xy,z)= b(x,z)b(y,z).
\end{align}
The first two equations imply that $q(x)$ is a well-defined function on $\pi_2=\ker \partial / {\rm im}\, \delta$ with values in $\pi_3 = \ker \delta $. The rest shows that $q(x)$ is a quadratic function.

Note that since 
\begin{align}
    \begin{CD}
 L @>u\circ g>> P.
\end{CD}
\end{align}
is a crossed module, for any $l$ such that $\delta l =(f(l^{-1}),g(l))= (1,1)$ we have
\begin{equation}
    ll'l^{-1}={}^{u(g( l))}l' = l',
\end{equation}
and thus $l$ belongs to the center of $L$. In particular, $\pi_3$ is abelian.

Similarly, one can show that   \begin{equation}
    Q /{\rm im\,} \delta \rightarrow P
\end{equation}
is a crossed module and $\pi_2$ is abelian.

% A word of caution: we work at the level of the non-abelian group $Q$ and use multiplicative notation. Once the above properties are proved, one can descend to Abelian $\pi_2$ and use additive notation for $q$ and $b$ as done in the main text.

In the following, all elements $x=(m,n),y=(m',n'),\dots$ of $Q = M\rtimes N$ are assumed to be closed, $\partial x = v(m)u(n)=1$, so $v(m)=u(n)^{-1}$.  
\subsection*{1. $\delta q(x)= 1$}
Writing the boundary operator explicitly, 
\begin{equation}
    \delta q(x) = \delta \eta(m,n)^{-1} = ( f(\eta(m,n)), g(\eta(m,n))^{-1} ).
\end{equation}
Using (\ref{eq: f eta}), we find
\begin{equation}
    f(\eta(m,n)) = m {}^{u(n)}m^{-1} =m {}^{v(m)^{-1}}m^{-1} =mm^{-1} m^{-1} m = 1,
\end{equation}
and similarly $ g(\eta(m,n))=1$.

\subsection*{2. $q(x \delta l)=q(\delta l x)=q(x)$}
First note that
\begin{equation}
    q(x \delta l) = \eta( m {}^{u(n)} f(l^{-1}), n g(l))^{-1}=\eta( m  f({}^{u(n)}l^{-1}), n g(l))^{-1},
\end{equation} 
where we used equivariance of $f$ with respect to action of $P$. Using (\ref{eq: eta exp}) repeatedly, we find
\begin{multline*}
    \eta( m  f({}^{u(n)}l^{-1}), n g(l)) \\= {}^{v(m)}\eta(  f({}^{u(n)}l^{-1}), n ) \eta(   f({}^{u(n)}l^{-1}), g(l))\eta( m  , n ){}^{u(n)}\eta( m  , g(l)).
\end{multline*}
Applying (\ref{eq: eta boundary}) simplifies the equation to
\begin{equation}
    \eta( m  f({}^{u(n)}l^{-1}), n g(l)) = {}^{u(n)}l l^{-1}\eta( m  , n )l{}^{u(n)}l^{-1},
\end{equation}
and since $\eta( m  , n )$ belongs to the center of $L$, we find
\begin{equation}
    q(x \delta l) = q(x).
\end{equation}
Due to image of $\delta$ being a normal subgroup of $Q$, the other equality
\begin{equation}
    q(\delta l x ) = q(x)
\end{equation}
follows. 
\subsection*{3. $q(x^n) = q(x)^{n^2}$}
For any $n'$, we have
\begin{equation}\label{eq: eta inv}
    {}^{u(n')} \eta(m,n) = \eta(m,n) \eta(f(\eta(m,n))^{-1},n')=\eta(m,n)\eta(1,n')=\eta(m,n),
\end{equation}
where we used (\ref{eq: eta boundary}), $\delta q(x)=1$, and the fact that $\eta(1,n)=\eta(m,1)=1$ which is a straightforward consequence of (\ref{eq: eta exp}). Similarly, $ {}^{v(m')} \eta(m,n)=\eta(m,n) $.

Now, repeatedly applying (\ref{eq: eta exp}) to $q(x^n)$, we arrive at $n^2$ terms $\eta(m,n)$ since all actions by images of $u$ and $v$ are trivial. Thus
\begin{equation}
    q(x^n)=q(x)^{n^2}.
\end{equation}
\subsection*{4. $b(x,yz)= b(x,y)b(x,z)$}

First note that for $x=(m,n)$, $y=(m',n')$ we have
\begin{equation}
    xy = (m {}^{u(n)} m', nn')= (m {}^{v(m^{-1})} m', nn')=(m'm, nn').
\end{equation}
Expanding $q(xy)$, we find
\begin{multline*}
    \eta(mm',n'n)= {}^{v(m)}\eta(m',n') {}^{v(m)u(n')}\eta(m',n)\eta(m,n'){}^{u(n')}\eta(m,n)\\=\eta(m',n') \eta(m',n'nn'^{-1})\eta(m,nn'n^{-1})\eta(m,n),
\end{multline*}
where we used invariance (\ref{eq: eta inv}), equivariance (\ref{eq: eta equivariance}), and 
\begin{equation}\label{eq: b invariance}
    {}^{v(m'')} l =\eta(m'',g(l)) l =\eta(m'',1)  l = l,
\end{equation}
where $m''$ is arbitrary, $l=\eta(m',n'nn'^{-1})\eta(m,nn'n^{-1})$ and we used (\ref{eq: f eta}, \ref{eq: eta boundary}). In particular, the last equation with $m''=f(l')$ means that  $\eta(m',n'nn'^{-1})\eta(m,nn'n^{-1})$ is central in $L$.

After this, $b(x,y)$ has a simple form 
\begin{equation}
    b(x,y) = \{x,y\}\{y,x\}
\end{equation}
(Recall that $\{x,y\}=\eta(m,nn'n^{-1})^{-1}$).

A long computation gives
\begin{multline*}
    b(x,yz)^{-1} =    {}^{u(n'n'')}\eta(m''m',n) {}^{u(n)}\eta(m,n'n'')\\ = {}^{u(n')}\eta(m',n) {}^{u(n'n'')}\eta(m'',n) {}^{u(n)}\eta(m,n'){}^{u(nn')}\eta(m,n'') \\= {}^{u(n')}\eta(m',n) {}^{u(n'n'')}\eta(m'',n) {}^{u(n'n)}\eta(m,n''){}^{u(n)}\eta(m,n')\\= {}^{u(n')}\eta(m',n) {}^{u(n'')}\eta(m'',n) {}^{u(n)}\eta(m,n''){}^{u(n)}\eta(m,n')\\={}^{u(n')}\eta(m',n) {}^{u(n)}\eta(m,n') {}^{u(n'')}\eta(m'',n) {}^{u(n)}\eta(m,n'') = b(x,y)^{-1}b(x,z)^{-1},
\end{multline*}
where we used
\begin{multline*}
 {}^{u(n)}\eta(m,n'){}^{u(nn')}\eta(m,n'') = {}^{ u(g({}^{u(n)}\eta(m,n')))u(nn')}\eta(m,n''){}^{u(n)}\eta(m,n') \\={}^{ u([n',n])u(nn')}\eta(m,n''){}^{u(n)}\eta(m,n') ={}^{ u(n'n)}\eta(m,n''){}^{u(n)}\eta(m,n'),
\end{multline*}
as well as invariance (\ref{eq: b invariance}) of $b(x,y)$ and the fact that it is central. The other identity $b(xy,z)= b(x,y)b(x,z)$ can be proved in the same way.

\bibliographystyle{unsrt} 
\bibliography{Bibliography.bib}

\end{document}